\documentclass[aps,prd,showpacs,12pt]{revtex4-1}
\usepackage{graphicx}
\usepackage{amsmath}
\begin{document}
\def\la{{\langle}}
\def\ra{{\rangle}}
\def\vep{{\varepsilon}}
\newcommand{\beq}{\begin{equation}}
\newcommand{\eeq}{\end{equation}}
\newcommand{\beqa}{\begin{eqnarray}}
\newcommand{\eeqa}{\end{eqnarray}}
\newcommand{\q}{\quad}
\newcommand{\AC}{{\it AC }}
\newcommand{\La}{{\Lambda }}
\newcommand{\lm}{{\lambda }}
\newcommand{\n}{\\ \nonumber}
\newcommand{\om}{\omega}
\newcommand{\e}{\epsilon}
\newcommand{\A}{\mbox A}
\newcommand{\Imm}{\mbox Im}
\newcommand{\Ree}{\mbox Re}
\newcommand{\B}{\mbox B}
\newcommand{\C}{\mbox C}
\newcommand{\D}{\mbox D}
\newcommand{\E}{\mbox E}
\newcommand{\Om}{\Omega}
\newcommand{\os}[1]{#1_{\hbox{\scriptsize {osc}}}}
\newcommand{\cn}[1]{#1_{\hbox{\scriptsize{con}}}}
\newcommand{\sy}[1]{#1_{\hbox{\scriptsize{sys}}}}
\newcommand{\pd }{Pad\'{e} }
\newcommand{\PAD }{Pad\'{e}\q}
\newcommand{\PD }{Pad\'{e}\hspace{2 mm}}
\newcommand{\get }{\leftarrow}
\newcommand{\f}{\ref }
\newcommand{\ics}{\texttt{ICS\_Regge} }

\title{Complex angular momentum theory of state-to state integral cross sections: resonance effects in the F+HD$\to$ HF($v'$=3)+D reaction\footnote{Dedicated to Prof. Tullio Regge (1931-2014)}}
\author {D. Sokolovski$^{a,b}$}
\author {E. Akhmatskaya$^{b,c}$}
\author {C. Echeverra-Arrondo$^a$}
\author {D. De Fazio$^{d}$}
\affiliation{$^a$ Departmento de Qu\'imica-F\'isica, Universidad del Pa\' is Vasco, UPV/EHU, Leioa, Spain,}
\affiliation{
$^b$ IKERBASQUE, Basque Foundation for Science, Maria Diaz de Haro 3, 48013, Bilbao, Bizkaia, Spain,}
\affiliation{$^c$ Basque Center for Applied Mathematics (BCAM), 
Alameda de Mazarredo, 14 48009 Bilbao, Bizkaia, Spain,
}
\affiliation{
$^d$ Istituto di Struttura della Materia, CNR, 00016 Roma, Italy}
\begin{abstract}
{\noindent
\begin{footnotesize}
State-to-state reactive integral cross sections (ICS) are often affected by quantum mechanical resonances, especially at relatively low energies. An ICS is usually obtained by summing partial waves at a given value of energy. For this reason, the knowledge of pole positions and residues in the complex energy plane is not sufficient for a quantitative description of the patterns produced by a resonance. Such description is available in terms of the poles of an $S$-matrix element  in the complex plane of the total angular momentum. The approach was recently implemented in a computer code \ics, available in the public domain [{\it Comp. Phys. Comm.} {\bf 185} (2014) 2127].
 In this paper, we employ  the \ics package to analyse in detail, for the first time, the resonance patterns predicted for integral cross sections (ICS) of the benchmark F+HD$\to$ HF($v'$=3)+D reaction. The $v=0$, $j=0$, $\Om=0$ $\to$ $v'=3$, $j'=0,1,2$, and $\Om'=0,1,2$ transitions are studied for collision energies from $ 58.54$ to $197.54$ $meV$. For these energies, we find several resonances, whose contributions to the ICS vary from symmetric and asymmetric Fano shapes to smooth sinusoidal Regge oscillations. Complex energies of metastable states and Regge pole positions and residues are found by \PD reconstruction of the scattering matrix elements. Accuracy of the  \ics code, relation between complex energies and Regge poles, various types of Regge trajectories, and the origin of the $J$-shifting approximation are also discussed.
 \end{footnotesize}}
\end{abstract}
\maketitle
{34.10,+x  34.50 Lf, }            
\section{introduction}
Recently there has been much interest in chemical reactivity in a cold environment, largely because of the spectacular progress in production and trapping of translationally cold molecules using a wide variety of experimental methods \cite{Krems09}. At low temperatures, reaction rates are often strongly influenced by quantum mechanical resonances and tunneling, which may cause large deviations of the rate constants from the Arrhenius law \cite{JPC14} -\cite{AR2}. Modelling of resonance phenomena, in which the reactants form intermediate complex before breaking up into products, requires large scale dynamical calculations on highly accurate potential surfaces. No less important is the task of interpreting the resonance patterns observed in the differential (DCS) and integral (ICS) cross sections. The DCS are more sensitive to the presence of rotating metastable complexes, which lead to observable nearside-farside oscillations in the angular distributions. A recent discussion of the interference effects in reactive DCS can be found, for example, in Refs. \cite{DCS1}-\cite{DCS2}.

Integral cross sections can also be affected by resonances. A resonance manifests itself in two types of singularities of a scattering matrix element ${\bf S}(E,J)$, which depends on two conserved quantities, the energy $E$ and the total angular momentum $J$. (We use bold symbols for complex valued quantities.) A pole of ${\bf S}$ as function of ${\bf E}$ for a fixed $J$ determines the energy and the lifetime of the metastable state. (In the following we will use bold symbols when we refer to complex mathematical quantities, and use normal characters otherwise.) In a similar way, a pole of ${\bf S}$ in the complex ${\bf J}$-plane at a fixed $E$ determines the angular momentum and the mean angle of rotation before the decay (lifeangle) of the complex formed at this energy.  
These complex angular momentum (CAM) or Regge poles \cite{REG1}-\cite{REG2}, are more useful for quantifying the resonance effects in observables, such as the ICS, which are  given by sums over partial waves at a given energy. This was first realised by Macek and co-workers, who related low-energy oscillations in the ICS for single-channel proton scattering by hydrogen atoms \cite{Macek1} and inert gases \cite{Macek2}, and gave a simple expression (known as the Mulholland formula \cite{Mull}) for the resonance contribution to an ICS. Their approach has proven to be useful for the studies of chemical reactions where resonances abound, and in Ref. \cite{JCP2} the Mulholland formula was extended to the multichannel case. 

For a single- or few-channel scattering problem, Regge poles can be found numerically by integrating the Schroedinger equation for complex-valued ${\bf J}$'s \cite{SE1,REG2,Macek1, SE2,INT2}, or by semiclassical methods \cite{Macek2}. This is no longer the case for a realistic reactive system with dozens of open channels, where evaluating the scattering matrix for physical integer values of $J$ is already a challenging task. Alternatively, it is possible to use the computed values of the ${\bf S}$-matrix elements in order to construct a \PD approximant, which will analytically continue $\bf S$ into the complex ${\bf J}$-plane \cite{PADE1}-\cite{PADE2}, and allow us to find the relevant Regge pole positions and residues. 
The method was implemented in the Fortran code $\texttt{PADE\_II}$ described in \cite{PADE2}, to which we refer the reader for more technical details. More recently, a code \ics designed specifically for analysing the ICS in a multichannel problem was reported in \cite{CPC2, SPRING}. The code, which uses the \PD reconstruction just outlined, identifies the trajectory, described by a Regge pole as energy varies, and evaluates its resonance contribution using the Mulholland formula of Ref.\cite{JCP2}. This allows us to decompose an often complicated pattern in the ICS into contributions from several Regge trajectories, corresponding to known resonance states. In this work, we use the \ics code in order to perform this analysis on the F+HD$\to$ HF($v'$=3)+D reaction at collision energies between $ 58.54$ and $197.54$ $meV$.

The F+HD$\to$ HF+D reaction (together with its isotopic variant F+H$_2\to$ HF+H) has earned its reputation as a benchmark case for the study of chemical kinetics after the 1985 molecular beam experiment of Ref. \cite{EXP1}, followed by molecular beam studies of higher resolution \cite{EXP2}-\cite{EXP9}. Most of these works provide evidence of the main resonance of the system (the resonance $\A$ in this work) affecting, in different ways, the total \cite{EXP2} and the state-to-state ICS \cite{EXP7},
 as well as the rotational distributions \cite{EXP9}, and the DCS \cite{EXP2,EXP3} of the HF($v'$=2) product. However, the presence of sharp forward and backward peaks in the DCS for the $v'=3$ vibrational manifold \cite{EXP2,EXP3}, suggested that an important role may be played by one or several resonance mechanisms also for this vibrational channel. 
Availability of  detailed experimental data prompted theoretical interest in the F+HD system, further encouraged by the fact that the {\it ab initio} potential energy surface (PES) obtained by Stark and Werner (SW) \cite{PES1} was able to reproduce, at least qualitatively, some of the resonance features observed \cite{EXP2}. The SW PES was later modified in the entrance channel region \cite{PES2}, in order to take into account the spin-orbit and the long-range interactions \cite{IJQC01}, in the light of previous elastic and inelastic molecular beam experiments \cite{Candori}. This surface, called PES III \cite{PES3}, was able to achieve a better agreement \cite{JCP06} in the height and shape of the resonance peak observed in \cite{EXP2}, as well as  predict quantitatively the cold (few Kelvin) behaviour of the F + H$_2$ reaction, recently measured in \cite{S57}. Dynamical calculations for the F+HD system on the PES III are reported, for example, in \cite{JCP08,JPC09,D1}. Analysis of \cite{JCP08,JPC09,D1} suggested a need for modifying
the PES also in the exit channel region, in order to achieve better agreement with experimental data, and, in particular, a better 
description of various resonance features.
Further progress  in this direction was made by Fu, Xu and Zhang (FXZ) \cite{PES4} obtaining a new {\it ab initio} PES able to reproduce with an impressive accuracy \cite{EXP7,D2} the resonance features observed. 
Moreover, the analysis reported in \cite{D2,JCP13} suggests a rich resonance pattern for the $v'$=3 manifold also predicted in the PES III calculations \cite{D1} and not yet fully explored in the experiments. 
Minor inaccuracies in the entrance and transition state regions of the PES, leading to small discrepancies  with experimental data, especially in the DF reactive channel \cite{D2}, have been  partially corrected recently in \cite{S63}.
In this work, we will use numerical S-matrix elements obtained in the dynamical calculations on the FXZ PES performed by De Fazio and co-workers \cite{D2}, as the input data for our analysis of resonance effects in the title reaction. 

The purpose of this paper is, therefore, to perform a detailed analysis of the effects produced by quantum mechanical resonances in the HF($v'$=3) ICS's of the benchmark F+HD reaction. We will also check the efficiency and accuracy of numerical techniques employed by the \ics.

The rest of the paper is organised as follows. In Section II we introduce the ICS's for the transitions of interest at collision energies close to the reactive thresholds and in the higher energy region, and obtain complex energies of the relevant metastable states as functions of the total angular momentum. In Sect. III we describe the Mulholland formula of Ref.\cite{JCP2}, and discuss two of its limiting cases. In Sect. IV we show that in the higher energy region the ICS's are affected by Regge oscillations produced by a single resonance. In Section V we discuss, in the simplest case, the relation between the complex energy of a metastable state and the corresponding Regge trajectory. In Sect. VI we analyse the behaviour of the ICS's in the neighbourhood of the reactive thresholds, where they are affected by several resonances. Section VII contains our conclusions.

\section{The integral cross sections}

A state-to-state reactive ICS is given by a partial wave sum (PWS) over the integer values of the total angular momentum $J$ (we are using $\hbar=1$)
 \begin{eqnarray}\label{b1}
 \sigma_{\nu^{\prime} \get \nu}(E) = 
 \frac{2\pi}{k_{\nu}^{2}}\sum_{J=J_{min}}^{\infty}(J+1/2)
 |{\bf S}_{\nu' \get \nu}(E,J)|^2,
\end{eqnarray}
where ${\bf S}_{\nu' \get \nu}(E,J)$ is the ${\bf S}$-matrix element for the reactants, prepared in the quantum state $\nu$, to form products in the state $\nu'$. For an atom-diatom collision, multi-index $\nu=(v,j,\Omega)$ includes the vibrational and rotational  quantum numbers of the diatomic, $v$ and $j$, and the absolute value of the diatomic's angular momentum projection, $\Omega$, onto atom-diatom relative velocity, also known as the helicity. Since this projection cannot exceed $J$, the summation in Eq.(\ref{b1}) starts from $J_{min}=max(\Omega,\Omega')$. Finally, $k_\nu$ is the reactant's wave vector. 

In this work we will analyse the resonance effects in the reactive transitions $(3,j'=0,1,2,\Om')\gets (0,0,0)$ in the interval of collision energies $ 58.54$ $meV  \le  E \le 197.54$ $meV$. Somewhat artificially, we divided this interval into 'lower' and 'higher' energy ranges, comprising $ 58.54$ $meV  \le  E \le 78.44$ $meV$, and $ 78.54$ $meV  \le  E \le 197.54$  $meV$, respectively. The reason for such a division becomes obvious from Fig.1 showing the ICS in both regions (note the change of the energy scale from Fig.1a to Fig1b). In the higher range shown in Fig.1b, all six ICS's exhibit similar slowly varying oscillatory structures superimposed on a smooth background. In the lower range shown in Fig.1a, the structure of the ICS's is much more complicated, with sharp peaks and saw-like features. This difference suggests that the two regions should be treated separately, as we will do in the following.
\begin{figure}
	\centering
		\includegraphics[width=15cm,height=10cm]{{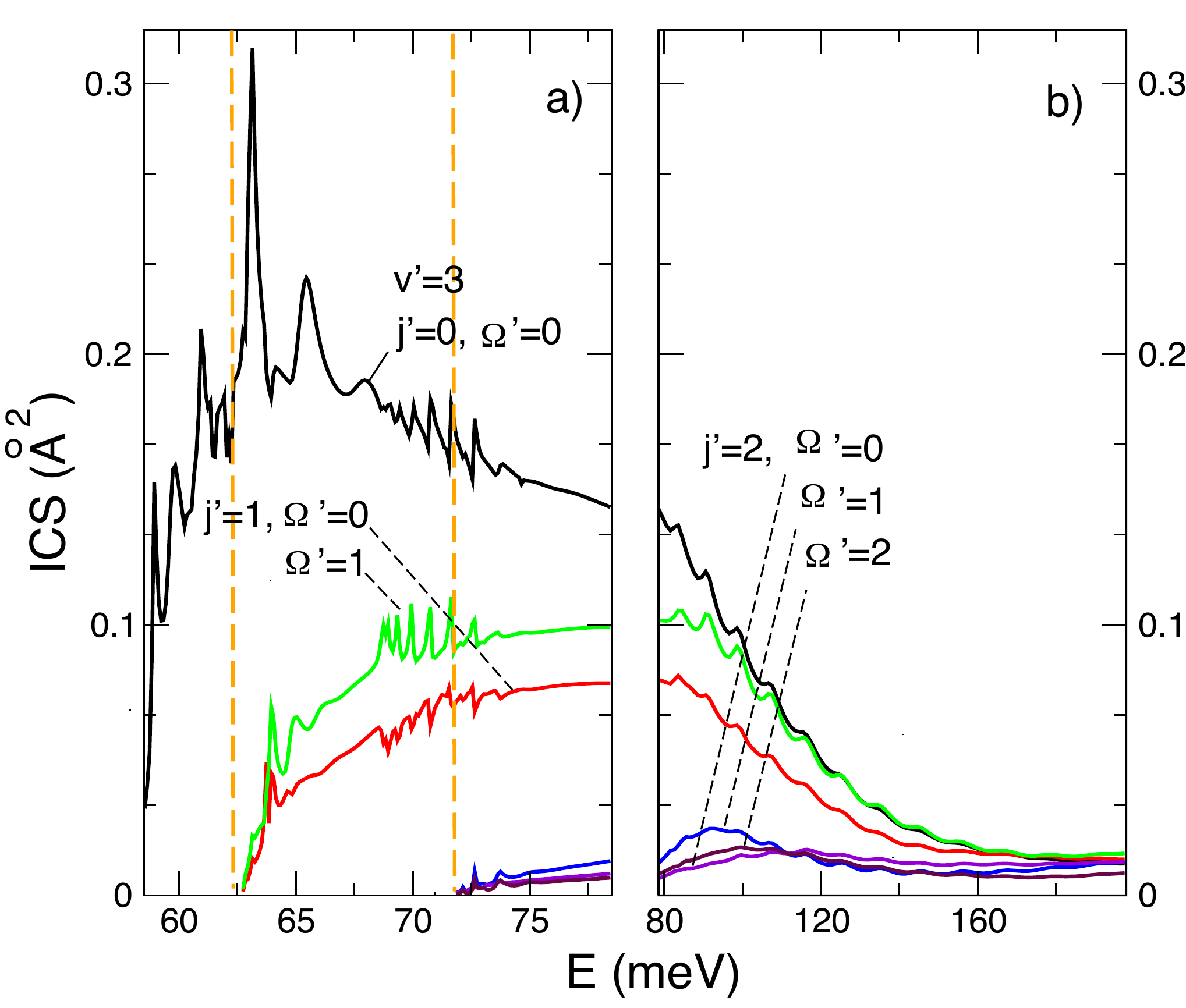}}
\caption{Six ICSs for the transitions $(v'=3,j'=0,1,2, \Om'=0,1,2)\gets(v=0,j=0,\Om= 0)$ in the lower (a) and higher (b) energy regions. Vertical dashed lines at $E=62.72$ and $71.77$ $meV$ indicate threshold energies beyond which the $j'=1$ and $j'=2$ transitions become possible. Note the difference in energy scales between the panels (a) and (b).}
\label{fig:1}
\end{figure}

\begin{figure}
	\centering
		\includegraphics[width=12.5cm,height=6cm]{{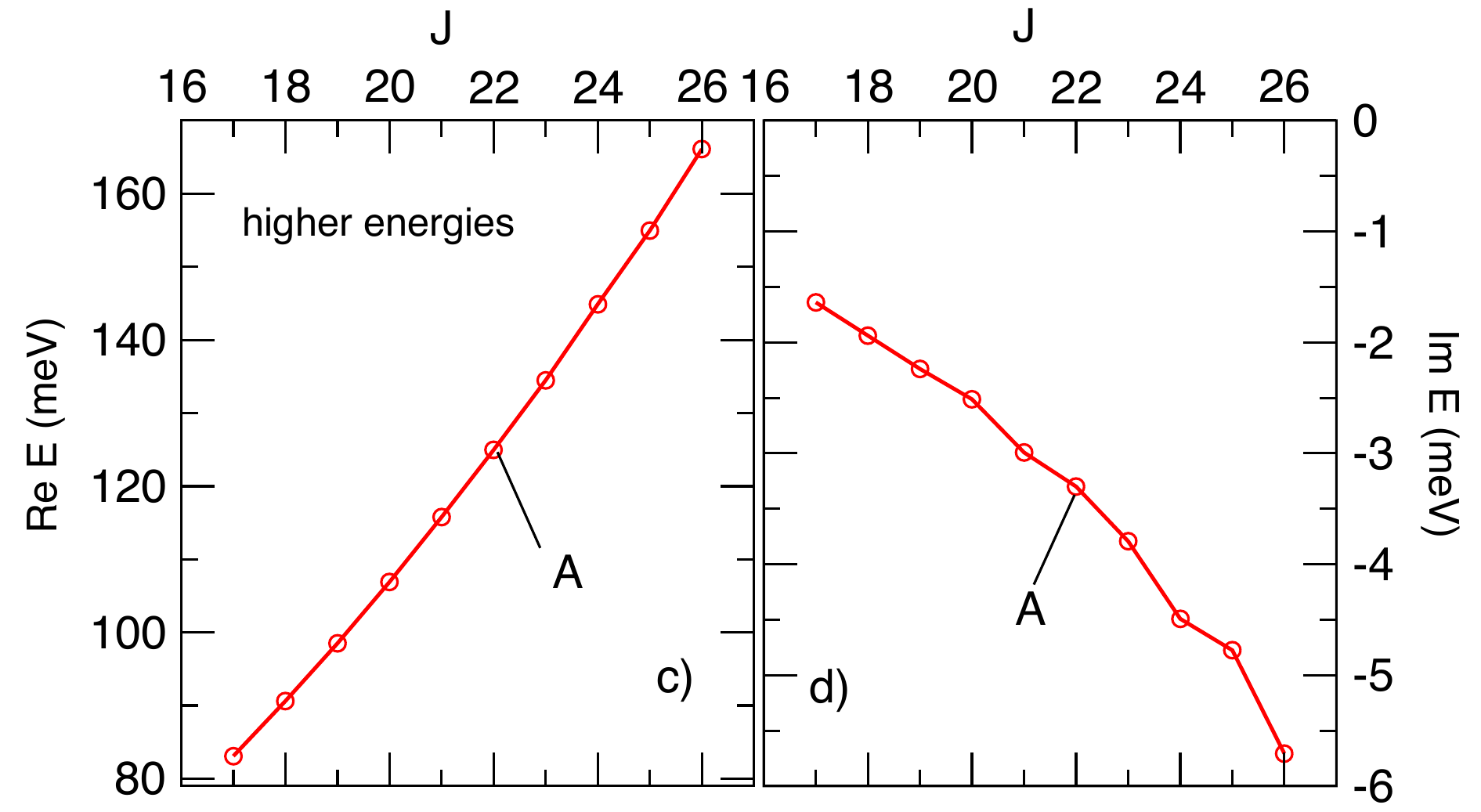}}
		\includegraphics[width=12cm,height=6cm]{{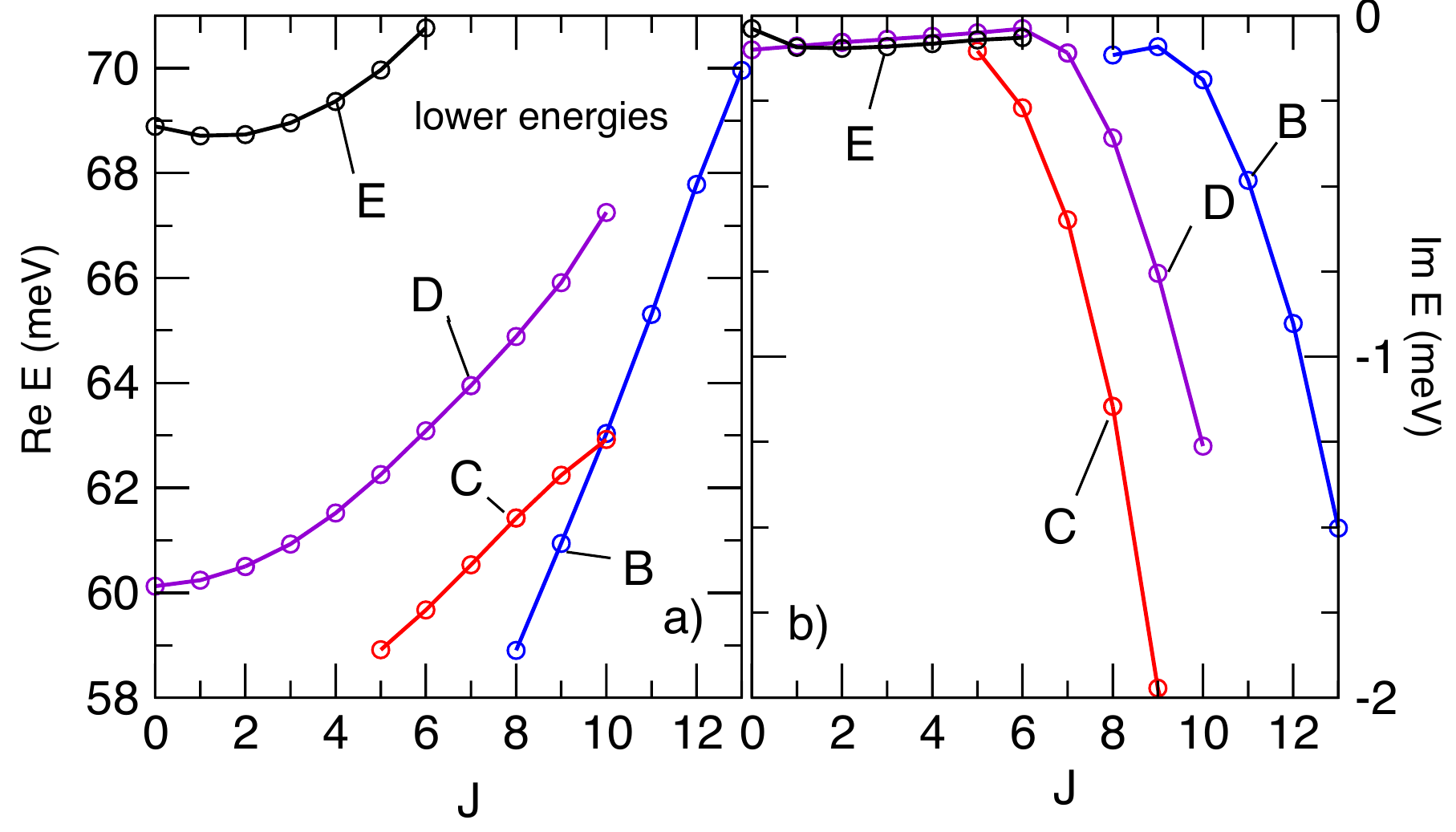}}
\caption{Complex energy poles found in the higher (c and d) and lower (a and b) energy ranges by \PD reconstruction of the $(3,0,0)\gets(0,0,0)$ ${\bf S}$-matrix element in the complex ${\bf E}$-plane for integer values of the total angular momentum $J$.}
\label{fig:2}
\end{figure}

As was mentioned in the Introduction, below we will use mostly the ${\bf S}$ matrix elements reported in Ref. \cite{D2}. However, to get an accuracy of about 1 $\%$ in the modules and phases of the single state-to-state ${\bf S}$-matrix elements, quantum reactive scattering calculations in the lower energy range were repeated using the improved code described in Refs. \cite{He12,DDF14}.
Also  for better convergence, the value of $\rho_{max} = 20$ $a.u.$ and $jmax = 25$ were used in these additional calculations. In particular, we will analyse in detail the ICS's for three reactive transitions, 
 \begin{eqnarray}\label{b1a}
 (3,0,0) \gets (0,0,0),\n
 (3,1,1) \gets (0,0,0),\n
 (3,2,2) \gets (0,0,0),
 \end{eqnarray}
 each representative of its rotational manifold. The behaviour for other allowed helicity states, $\Om'=0$ for $j=1$, and $\Om'=0,1$ for $j=2$, is very similar, and would contribute little new information. 

\section{The complex energy poles}

We expect the structures in the ICS's in Fig. 1 to be signatures of a resonance mechanism, or mechanisms, and look for a further confirmation. One usually starts by looking  for poles of the ${\bf S}$-matrix element in the complex energy (CE)  plane, at physical values of the total angular momentum, $J=0,1,2....$. There is an ongoing discussion (see \cite{DCS2} and Refs. therein) about when a particular pole, or a combination of poles, should be considered a resonance. We will avoid the controversy by directly identifying a pole or poles, and quantifying their effect on the integral cross sections of interest. Accurate pole positions can be obtained, for instance, by evaluating Smith's lifetime ${\bf Q}$-matrix \cite{Smith} using  numerical differentiation of the ${\bf S}$-matrix elements \cite{DARIO04}, or by a direct approach described in Ref. \cite{JCP05}.  Here we use a different method, based on \pd reconstruction of the chosen ${\bf S}$-matrix element. The \pd reconstruction analytically continues ${\bf S}_{\nu' \get \nu}(E,J)$, evaluated for a set of real energies, into the complex ${\bf E}$-plane (see, for example, \cite{PADE2}).

Plotting the CE pole positions for different values of $J$ we obtain CE trajectories, shown in Fig. 2 for the $(3,0,0)\leftarrow(0,0,0)$ reactive transition.  We note that there is only one CE pole in the higher energy region, while in the lower energy range there are many CE trajectories. While a CE pole at ${\bf E}_0$   is the same for all the elements of the ${\bf S}$-matrix \cite{Land}, the extent to which it affects the reaction probability $P_{\nu' \get \nu}(E,J)=|{\bf S}_{\nu' \get \nu}(E,J)|^2$ depends on the corresponding residue, $\boldsymbol \rho_{\bf E}=\mbox{lim}_{{\bf E}\to {\bf E}_0}({\bf E}-{\bf E}_0){\bf S}_{\nu' \get \nu}({\bf E},J)$, specific to each transition. Thus, certain transitions may be affected by a particular CE pole, while some may not, and a different choice of $\nu$ and $\nu'$ may result in different sets of trajectories.

The CE trajectories for the F+HD reaction have been studied earlier \cite{D1,D2}. In \cite{D1} the authors have obtained such trajectories for a different PES, PES-III \cite{PES3}, by fitting the peaks in the highest eigenvalue of the ${\bf Q}$-matrix to a Lorentzian form. A preliminary study for the PES used in this work (the FXZ PES \cite{PES4}) was reported in \cite{D2}, where the authors followed some of the peaks in the reaction probability, thus approximating only the real parts of the corresponding CE trajectories, with the results similar to those shown in Figs.  2 a and b. The five poles shown in Fig. 2 were labelled in Refs. \cite{D1,D2,DARIO04,JCP05}, as $\A$, $\B$, $\C$, $\D$ and $\E$, a notation originally introduced by Manolopoulos and co-workers \cite{mano96}, who were the first  to discuss their effect on the J=0 reaction probability of the F+H$_2$ system. We will continue to use the same nomenclature in what follows. Analysing the build up of the scattering wave function in a particular spacial region \cite{D1}, or studying the adiabatic curves \cite{DARIO04} often allows one to identify the physical origin of a particular CE pole. Thus, the pole $\A$ corresponds to trapping the reactants close to the transition state of the PES, while the poles $\B$, $\C$, $\D$, and $\E$ describe capture in the metastable states of the van der Waals well in the exit channel \cite{D1} corresponding to different bending and symmetric stretching modes \cite{JCP05}.

The complex energy poles shown in Fig. 2 may suggest that the patterns seen in Fig.1 are related to the resonances $\A$, $\B$, $\C$, $\D$ and $\E$. Complex energies, however, are not particularly useful for our task of quantifying the resonance effects produced in the ICS's. Indeed, poles are most useful when there is an integration and, by changing the contour and applying the Cauchy theorem, one can isolate a pole's contribution. Thus, the CE poles would be useful, if we were to evaluate integrals over energy.  In Eq.(\ref{b1}), however, $E$ is a fixed parameter, and the summation is over $J$'s. Sums over $J$ can be transformed into integrals, e.g., by applying the Poisson sum formula, and to evaluate such integrals we would need the other kind of poles.

\section{Regge pole decomposition of an integral cross section}
         
The poles we need are the poles of the ${\bf S}$-matrix element in the complex ${\bf J}$-plane at a fixed real energy, also known as complex angular momentum (CAM) or Regge poles. The elements of the scattering matrix diverge whenever the solution of the corresponding Schroedinger equation contains only outgoing waves at large interatomic distances. At a given real energy, this can only happen in special cases, when the complex valued centrifugal potential is of emitting kind.
Thus, the positions in the first quadrant of the complex ${\bf J}$-plane, are given by the values ${\bf J}_n$, $n=1,2,..$ for which ${\bf S}_{\nu' \get \nu}(E,{\bf J}_n)$ becomes infinite. Following \cite{Macek1,JCP2,CPC2,SPRING} we rewrite Eq.(\ref{b1}) in a different form, separating the resonance contributions from the slowly varying 'background' part,
\begin{eqnarray}\label{a11}
 \sigma_{\nu^{\prime} \gets \nu}(E) = \sum_n \sigma_n^{res}(E)+\sigma^{back}(E),
\end{eqnarray}
where the summation is over all resonance poles. Each pole, e.g., the pole number $n$ whose position is ${\bf J}_n$, contributes to the ICS 
\begin{eqnarray}\label{a12}
 \sigma_n^{res}(E)=\frac{8\pi^2}{k_\nu^2}\Imm\frac{\boldsymbol \lambda_n \boldsymbol \rho_n(E){\bf S}^{*}_{\nu' \leftarrow \nu}(E,\boldsymbol \lambda_n^*)}{1+\exp(-2i\pi \boldsymbol \lambda_n)},
\end{eqnarray}
where 
$$\boldsymbol \lambda_n \equiv {\bf J}_n+1/2,$$
$\boldsymbol \rho_n$ is the residue, 
\begin{eqnarray}\label{a13}
\boldsymbol \rho_n(E)=lim_{{\bf J}\to {\bf J}_n}(\boldsymbol \lambda-\boldsymbol \lambda_n){\bf S}_{\nu' \leftarrow \nu}(E,\boldsymbol \lambda),
\end{eqnarray}
and ${\bf S}^{*}_{\nu' \leftarrow \nu}(E,\boldsymbol \lambda_n^*)$ the complex conjugate of ${\bf S}_{\nu' \get \nu}$, evaluated at the point conjugate to the poles own position, ${\bf J}_n^*$.

The background part, $\sigma^{back}(E)$, consists of two terms ($\lambda\equiv J+1$)
\begin{eqnarray}\label{a14}
\sigma^{back}(E) = \frac{2\pi}{k_\nu^2}\int_{J_{min}}^\infty|{\bf S}_{\nu' \leftarrow \nu}(E,\lambda)|^2 \lambda d\lambda+I_{\nu' \leftarrow \nu}(E). \q\q\q
\end{eqnarray}
The first one is just what one would get by replacing the  sum in (\ref{b1}) by an integral over $J$. The second one contains integrals along the imaginary $\boldsymbol \lambda$-axis, and is expected to be both small and smooth. Normally there is no need to evaluate it explicitly, since, if needed, it can be obtained by subtracting $\sum_n \sigma_n^{res}(E)$ and the integral in Eq.(\ref{a14}) from the exact ICS (\ref{b1}) .

The usefulness of the Mulholland formula (\ref{a11}-\ref{a14}) can be illustrated by the following example. Like a CE pole, a CAM pole cannot vanish suddenly. Rather, it moves in a continuos manner as the energy changes, describing what is known as Regge trajectory \cite{SE1}. Consider a pole ${\bf J}_0(E)$ with a small residue and so close to the real ${\bf J}$-axis, that it makes to ${\bf S}_{\nu' \get \nu}(J)$  a contribution, whose width is much smaller than $1$. If at an energy $E$ the pole happens to lie close to an integer value, say, of $\Ree {\bf J}_0(E)\approx5$, the fifth partial wave in Eq.(\ref{b1}) will be enhanced, and the value of the ICS will increase as a result. Change the energy a little, so that the narrow shape now fits between $J=5$ and $J=6$, and neither of the two partial waves is enhanced. The value of the ICS will fall back, and will increase again as the pole passes near the next integer value of $\Ree {\bf J}_0(E)\approx6$.

The first term in Eq.(\ref{a14}) is an integral over $J$, rather than a sum. It will always receive a similar contribution  from the Regge pole, and, therefore, be a smooth function of energy. The rapid variations of the ICS are now contained in the first term in Eq.(\ref{a11}), which will be enhanced each time the pole passes near an integer, where its denominator has a minimum \cite{Macek1}. Thus, a Regge trajectory passing near the real ${\bf J}$-axis, $\Imm {\bf J}_n <<1$, produces in the ICS a series of sharp features, centred at the energies $E_n^{K}$ at which the trajectory is close to a real integer value of $J$, $\Ree {\bf J}_n(E_n^{K}) = K$, $K=0,1,2,...$ \cite{JCP2},
\begin{eqnarray}\label{a15}
 \sigma_n^{res}(E)\approx \frac{-2\pi}{k_\nu^2}\Ree\sum_K\frac{\boldsymbol \gamma_n^K}{E-E_n^K+i\Gamma^K_n/2}\equiv \sum_K \sigma_{n,K}^{res}(E),
\end{eqnarray}
where 
\begin{eqnarray}\label{a16}
 \boldsymbol \gamma_n^K\equiv \boldsymbol \lambda_n \boldsymbol \rho_n {\bf S}^{*}_{\nu' \leftarrow \nu}(E,\boldsymbol \lambda_n^*)/\partial_E \Ree \boldsymbol \lambda_n |_{E=E_n^K},
\end{eqnarray}
and the peak's width at the $K$-th energy, $\Gamma_n^K$, is given by
\begin{eqnarray}\label{a17}
 \Gamma_n^K\equiv  2\Imm \boldsymbol \lambda_n /\partial_E \Ree \boldsymbol \lambda_n |_{E=E_n^K}.
\end{eqnarray}
We note that $\sigma_{n,K}^{res}(E)$ may have either sign, and is not, in general symmetric around the energies $E_n^K$. Equation (\ref{a15}) is a variant of the celebrated Fano line shape formula \cite{FANO1,FANO2,FANO}. Like the Fano result, $\sigma_{n,K}^{res}(E)$ in Eq.(\ref{a15}) reduces to a Breit-Wigner Lorentzian shape for $\Ree [\boldsymbol \gamma_n^K] =0$, and is antisymmetric about $E=E^K_n$ for $\Imm [\boldsymbol \gamma_n^K] =0$.

 For a pole passing at a larger distance from the real ${\bf J}$-axis, the sequence of Fano-like features in the ICS is replaced by smoother sinusoidal oscillations whose maxima do not necessarily coincide with $E_n^K$, and we have \cite{JCP2}
 \begin{eqnarray}\label{a18}
 \sigma_n^{res}(E)\approx \frac{4\pi^2}{k_\nu^2}|\boldsymbol \lambda_n \boldsymbol \rho_n {\bf S}^{*}_{\nu' \leftarrow \nu}(E,\boldsymbol \lambda_n^*)|\n
 \times \exp(-2\pi \Imm \boldsymbol \lambda_n) \sin [2\pi \Ree \boldsymbol \lambda_n(E)+\phi(E)],
\end{eqnarray}
 where $\phi(E)$ is the phase of the product between the moduli signs,
  \begin{eqnarray}\label{a19}
\phi(E)\equiv \mbox {Arg}\{ \boldsymbol \lambda_n \boldsymbol \rho_n {\bf S}^{*}_{\nu' \leftarrow \nu}(E,\boldsymbol \lambda_n^*)\}.
\end{eqnarray}
Equation (\ref{a18}) helps illustrate a general trend: a resonance contribution increases with the increase of the residue $\boldsymbol \rho$, and is rapidly quenched, owing to the presence of $\exp(-2\pi \Imm \boldsymbol \lambda_n)$, if a trajectory moves deeper into the complex ${\bf J}$-plane.

Finally, for a pole further away from the real axis, $\Imm {\bf J}_n \gtrsim 1$, the resonance term will vanish, unless its residue is extremely large. Such pole may, however, continue to influence angular distributions, more sensitive to the presence of resonances (see, for example, \cite{G1}).

To use the Mulholland formula in Eq.(\ref{a11}) one needs to know the behaviour of the ${\bf S}$-matrix element in the complex ${\bf J}$-plane. For this purpose we construct a \pd approximant from the known values of ${\bf S}_{\nu' \get \nu}$  at integer values of $J$ using the package \ics \cite{CPC2}. The approximant analytically continues the ${\bf S}$ matrix element into the complex ${\bf J}$-plane, and correctly represents it in the vicinity of the integral $J$ values used in its construction. For resonance poles sufficiently close to the real ${\bf J}$-axis we can, therefore, evaluate all relevant quantities in Eqs.(\ref{a11})-(\ref{a14}). The poles further away from the real axis may not be represented correctly. But this does not matter, since the resonance contributions of such poles vanish due to the rapidly growing factor $\exp(-2i\pi\boldsymbol \lambda_n)$ in Eq.(\ref{a12}). For more information about \PD reconstruction of numerical scattering matrix we refer the reader to Refs. \cite{G1} and \cite{PADE2}.

\section{The higher energy range:  the resonance A and its Regge oscillations}

\begin{figure}
	\centering
		\includegraphics[width=15cm,height=10cm]{{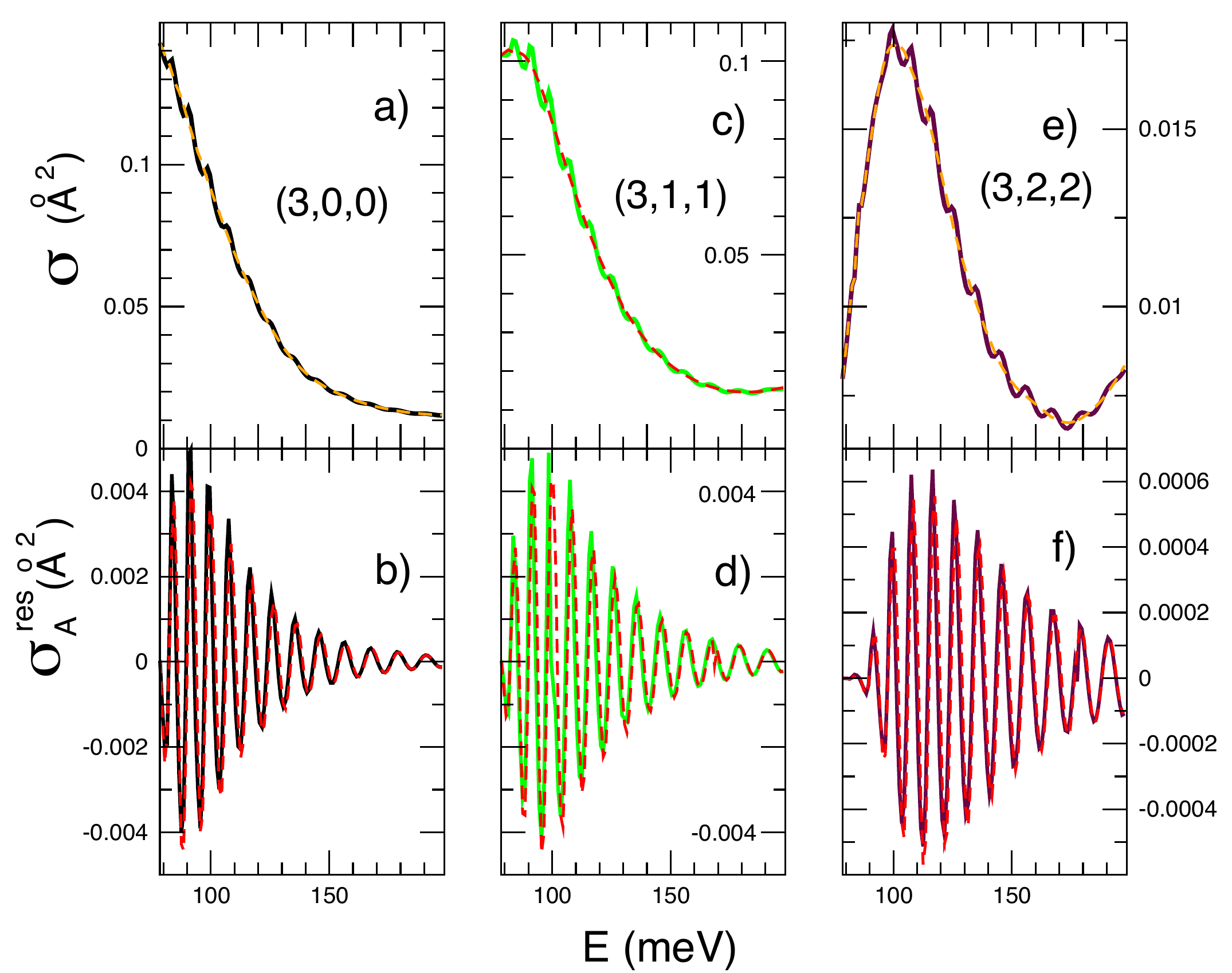}}
\caption{ a) The ICS for the $(3,0,0)\gets(0,0,0)$ reactive transition (solid). Also shown is the difference $\sigma(E)-\sigma^{res}(E)$ (dashed); c) same as a) but for the $(3,1,1)\gets(0,0,0)$ transition; e) same as a) but for the $(3,2,2)\gets(0,0,0)$ transition ; b) the resonance contribution $\sigma^{res}(E)$ to the transition in a), exact (solid) and given by Eq.(\ref{a18}) (dashed); d) same as b) but for the transition in c); f) same as b) but for the transition in e).}
\label{fig:3}
\end{figure}

Next we apply the analysis just described to the three reactive transitions in (\ref{b1a}) in the higher energy range $78.5$ $ meV < E <198.5$ $meV$. In this range, all three ICS's show oscillatory structures superimposed on a slowly varying background (see Figs. \ref{fig:3}a, \ref{fig:3}c, and \ref{fig:3}e). 

\begin{figure}
	\centering
		\includegraphics[width=10cm,height=10cm]{{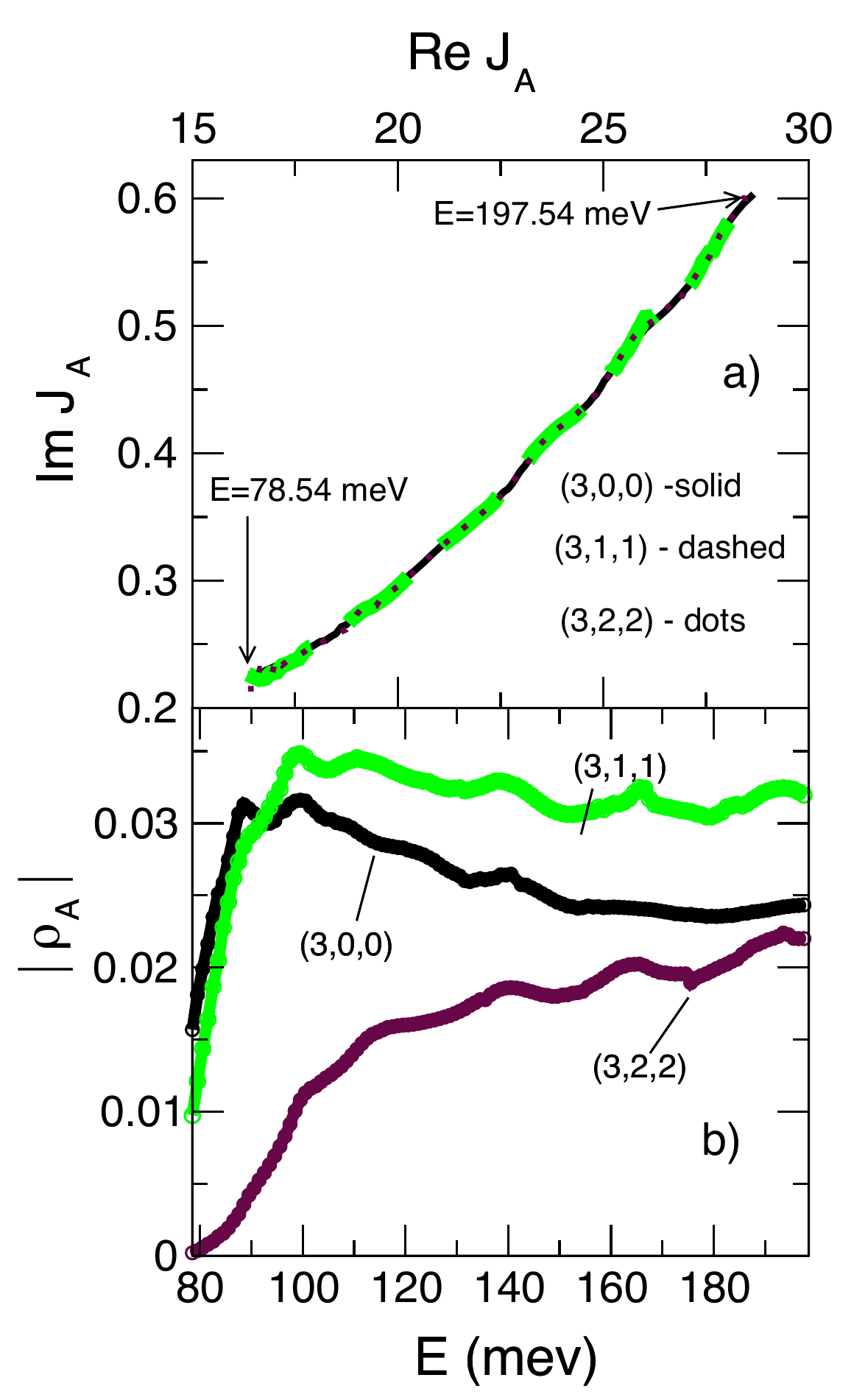}}
\caption{ a) Regge trajectories for the transitions in (\ref{b1a}) in the higher energy region;
b) residues for three trajectories shown in a).} 
\label{fig:4}
\end{figure}

With the help of the \ics program \cite{CPC2} we found three significant Regge trajectories in the chosen energy range, one for each transition. Both real and imaginary parts of the pole positions increase with energy, as the trajectory moves deeper into the complex ${\bf J}$-plane. We note that the trajectories obtained for each of the three transitions (\ref{b1a}), shown in  Fig. \ref{fig:4}a, coincide to almost graphical accuracy, indicating that all the transitions are affected by the same resonance. Like a CE pole \cite{Land}, a Regge pole affects, in principle, all elements of an ${\bf S}$-matrix. A degree to which an element is affected depends on its residue at the pole in question. For many transitions, the residue is zero, or extremely small, so  that only particular ro-vibrational manifolds (e.g., $v'=3$, $j'=0,1,2$ in this study) are affected by the same pole. Finding a serious discrepancy between the Regge trajectories obtained for the affected transitions, would indicate a defect of the numerical method used for their evaluation. Finding all three in good agreement is an indication that the method works correctly. 

 Figure \ref{fig:4}b shows the behaviour of the three residues whose magnitudes, after an initial rapid increase tend to level off at higher energies.  The resonance contributions to the ICS's, $ \sigma^{res}(E)$, shown in Figs.\ref{fig:3}b, \ref{fig:3}d and \ref{fig:3}f, exhibit modulated Regge oscillations well described by Eq.(\ref{a18}). At lower energies, they are limited by the magnitude of the corresponding residue, while at higher energies they vanish as the imaginary part of the pole position increases [cf. Eq.(\ref{a18})]. This is a behaviour typical for a Regge trajectory passing not too close to the real axis \cite{JCP2}. We note that Regge oscillations for different transitions are in phase and, for this reason, they remain visible in the reactive ICS summed over the rotational quantum numbers and helicities \cite{D2}. 

Comparison with Fig. \ref{fig:2} suggests that the resonance responsible for the patterns in Figs. \ref{fig:3}a, \ref{fig:3}c and \ref{fig:3}e is the one we earlier called $\A$. In the next Section we will provide a more rigorous proof.

\section{The relation between complex energies and Regge poles}

The poles of ${\boldsymbol S}$-matrix in the complex ${\boldsymbol E}$- and ${\bf J}$-planes, introduced in Sects. III and IV, are intimately related. An ${\bf S}$-matrix element is a function of two independent variables, $E$ and $J$, which enters the Schroedinger equation in the combination $\Lambda=J(J+1)$. Treating $E$ as a real or complex parameter, we recall that a pole at some ${\boldsymbol \La}_n={\bf J}_n({\bf J}_n+1)$, ${\bf S}_{\nu'\gets \nu}(E,{\bf J}_n)=\infty$, occurs if the solution of the Schroedinger equation, regular at the origin, contains only outgoing waves at infinity \cite{REG1}. Thus, ${\boldsymbol \La}_n$ is a complex eigenvalue of a Sturm-Liouville problem \cite{IH1, Macek1}. Typically, there are infinitely many such eigenvalues ${\boldsymbol \La}_n({\boldsymbol E})$, $n=1,2,...$, for each value of $E$, whether real or complex. Overall, ${\boldsymbol \La}({\boldsymbol E})$ can be seen as a single-valued function defined on a multi-sheet Riemann surface $\mathcal{R}$, whose sheets may be connected, e.g., at branching points. The $n$-th Regge trajectory, ${\bf J}_n(E)$, can then be found by moving along the real $E$-axis on the $n$-th Riemann sheet, and reading off the values of ${\boldsymbol \La}(E)$ as we go. Inverting ${\boldsymbol \La}({\boldsymbol E})$ we have a function ${\boldsymbol {\mathcal{E}}}(\boldsymbol\La)$, single-valued on its own Riemann surface $\mathcal{R'}$ which contains the information about the position of complex energy poles. Although neither ${\boldsymbol \La}({\boldsymbol E})$ nor $\mathcal{R}$ are usually known, the relation between CE and Regge poles can be explored locally, that is, for a single isolated trajectory  close to the real axis, and in limited ranges of ${\boldsymbol E}$ and ${\boldsymbol J}$. 

 Below we consider this simple case, suspending for a time the subscript $n$ numbering the pole. Suppose we have a segment of the real 
$\Lambda$-axis, $\Delta \Lambda$, where, for a CE trajectory ${\boldsymbol E}(\Lambda) = E_1(\La)+iE_2(\La)$, both $E_1$ and $E_2$ are approximately linear functions of $\La$. Since ${\bf \mathcal{E}}({\boldsymbol \La})$ is analytic, it  is, approximately, a linear function of ${\boldsymbol \La}$, ${\boldsymbol E}({\boldsymbol \Lambda}) \approx {\bf A}\La+{\bf B}$, in a subset of $\mathcal{R}$ around $\Delta \La$, which we will call $w$. The function ${\bf \mathcal{E}}(\La)$ transforms $w$ into a subset $w'$ of the complex ${\bf E}$-plain, where (we write ${\bf Z}=Z_1+iZ_2$)
\begin{eqnarray}\label{a2}
\left( \begin{array}{rr} E_1 \\ E_2 \end{array}\right) = \left( \begin{array}{rr} A_1 & -A_2 \\ A_2 & A_1 \end{array}\right)\left( \begin{array}{rr} \La_1 \\ \La_2 \end{array}\right)+\left( \begin{array}{rr} B_1 \\ B_2 \end{array}\right).
\end{eqnarray}
 If $E_2(\La)$ is sufficiently small, there is also a segment of the real ${\boldsymbol E}$-axis, $\Delta E$, which lies in $w'$. The image of $\Delta E$ in the complex ${\boldsymbol \La}$-plane can be obtained by inverting the linear relation (\ref{a2}) 
\begin{eqnarray}\label{a3}
\left( \begin{array}{rr}\La_1 \\\La_2 \end{array}\right) = \Delta^{-1}\left( \begin{array}{rr} A_1 & A_2 \\ -A_2 & A_1 \end{array}\right)\left( \begin{array}{rr} E_1 \\ E_2 \end{array}\right)- \Delta^{-1}\left( \begin{array}{rr} A_1B_1+A_2B_2 \\ A_1B_2-A_2B_1\end{array}\right),
\end{eqnarray}
where $\Delta=A_1^2+A_2^2$, and putting $E_2=0$. Expressing ${\boldsymbol J}$ as a function of ${\boldsymbol \La}({ E})$ yields the Regge trajectory ${\boldsymbol J}(E)$. Alternatively, we may be interested in how the position of a CE pole ${\bf E}_n$ varies with $J$, while $J$ is kept real, $\Imm J=0$. In this case we put  $\La_2=0$ in Eq.(\ref{a2}) and solve it for ${\boldsymbol E}(\La_1)$.  

Relating CE and CAM pole trajectories can be useful if, for example, the two trajectories have been obtained independently, and the consistency of the calculations needs to be checked. It may also be useful if there are many trajectories of each kind, and one needs to know which CAM trajectory corresponds to a particular CE one. Finally, the physical mechanism of a resonance, e.g., capture in the van der Waals well of the exit channel, is most easily established for CE poles \cite{D1}, \cite{D2}. The resonance contributions to observable cross sections, on the other hand, are best described with the help of CAM trajectories \cite{CPC2}. Again, the ability to relate the two types of trajectories can be helpful.

\begin{figure}
	\centering
		\includegraphics[width=9cm,height=6cm]{{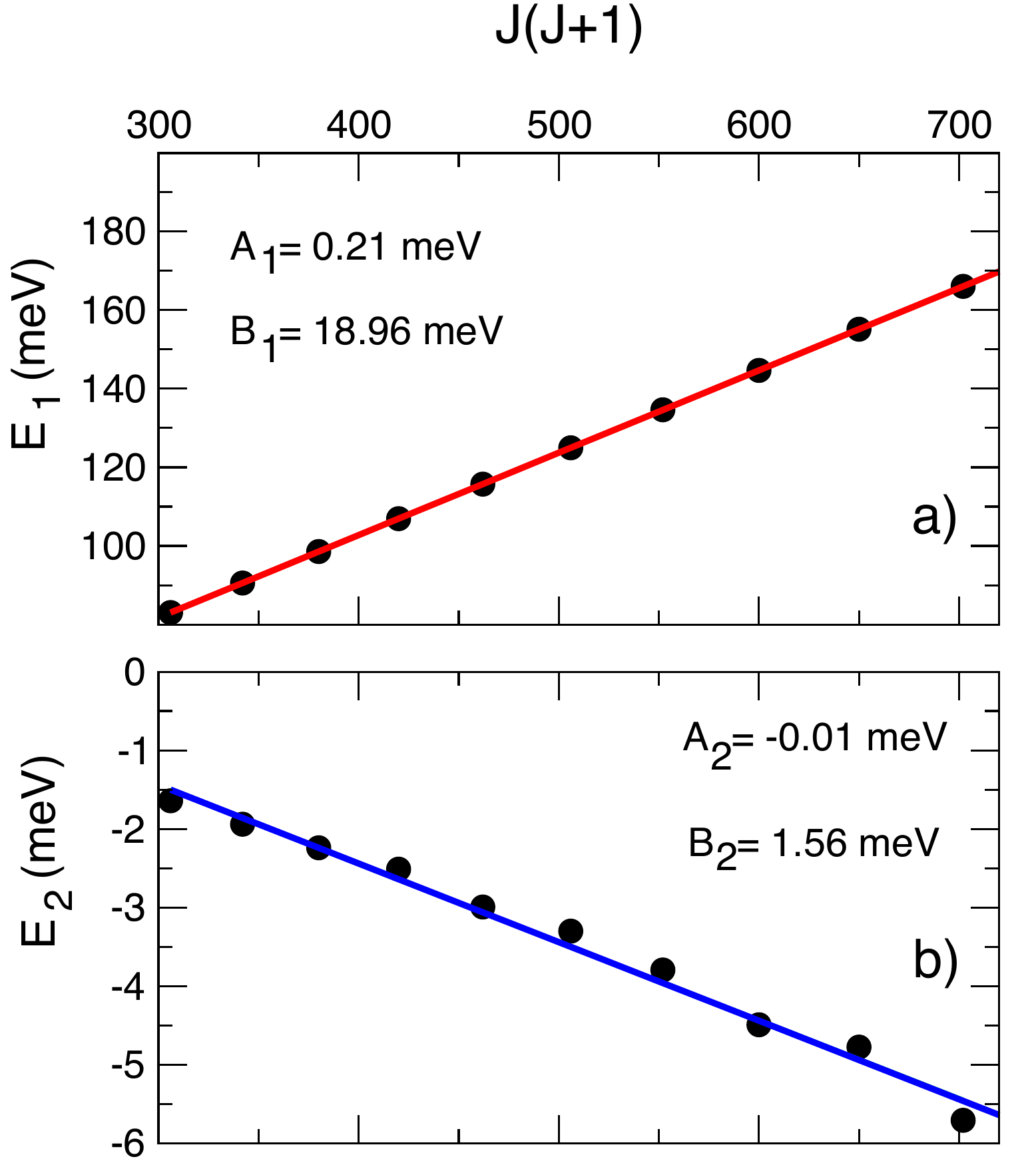}}
		\includegraphics[width=9cm,height=6cm]{{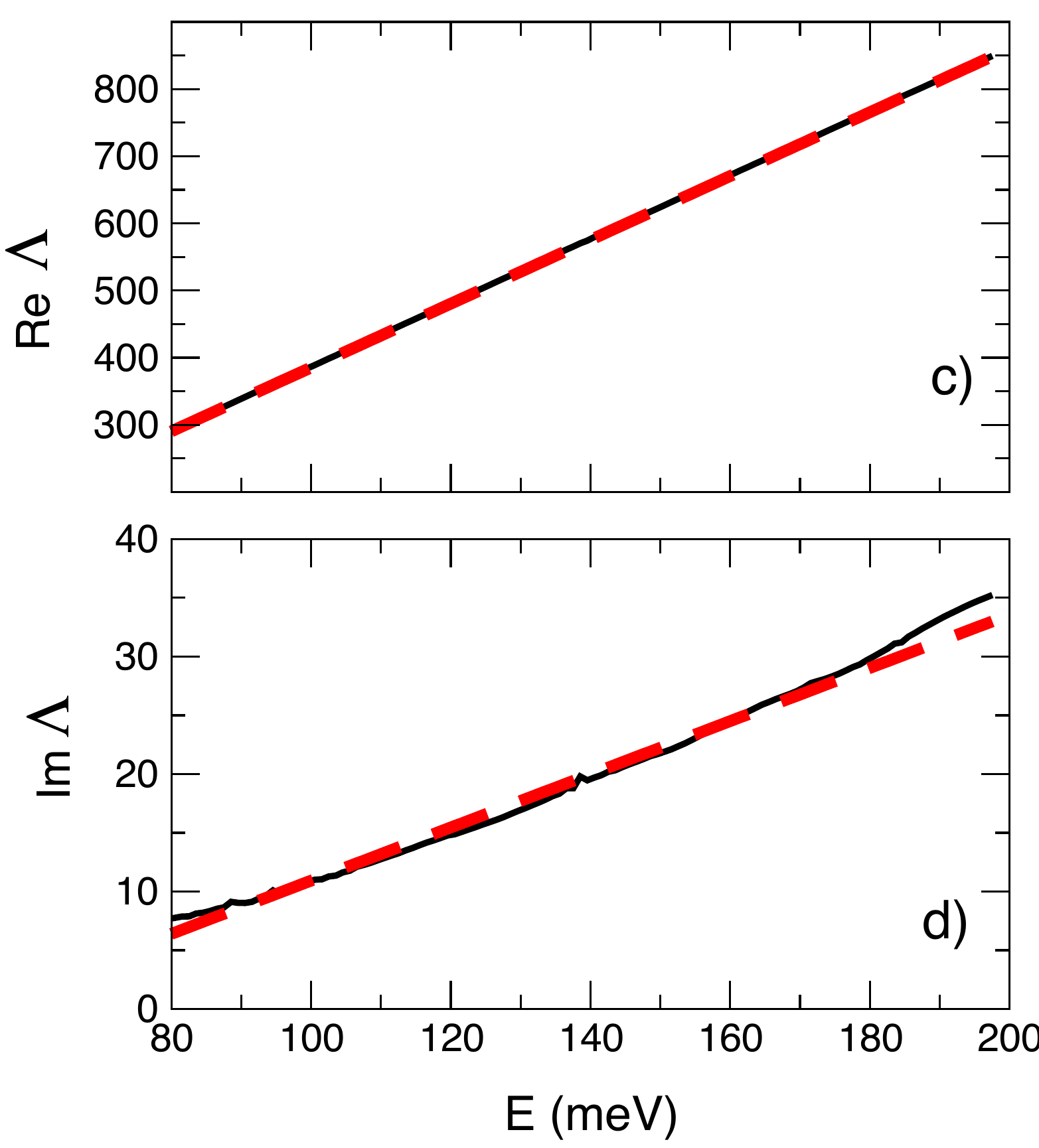}}
\caption{ a) Real part of the position of the pole A in the complex energy plane vs. $\Lambda=J(J+1)$ evaluated by \PD reconstruction of the ${\bf S}$-matrix element for real integer values of $J$,  $J=17,...,27$ (dots), and its linear fit corresponding to Eq.(\ref{a2}) (solid); b) same as (a), but for the imaginary part of the pole's position; c) real part of ${\boldsymbol \La}(E)$ evaluated for the pole A by \PD reconstruction of the ${\bf S}$-matrix element for $78.54\q meV < E <197.54\q meV$ (solid), and as predicted by Eq.(\ref{a3}) (dashed); d) same as (c), but for the imaginary part. 
}
\label{fig:5}
\end{figure}

 As an example, we check whether the Regge trajectory shown in Fig.\ref{fig:4}a for the $(3,0,0)\gets (0,0,0)$ transition is indeed the CAM counterpart of the CE trajectory in Fig.\ref{fig:2}, which we have earlier labelled $\A$. We do it by first evaluating the CE trajectory for the pole A in the higher energy region numerically. We then fit the pole position by a linear function of $\Lambda =J(J+1)$, thus obtaining the values for coefficients $A_{1,2}$ and $B_{1,2}$ in Eq.(\ref{a2}). These values are inserted into Eq.(\ref{a3}), and the approximate Regge trajectory for the pole $\A$ is drawn.  This trajectory is then compared with the exact Regge trajectory obtained independently, by \PD reconstruction of the ${\bf S}$-matrix element in the energy range of interest. A good agreement between the two CAM trajectories shown in Fig. \ref{fig:5} demonstrates the validity of our analysis.

It is instructive to revert the discussion to more physical terms, in the simple case when the imaginary part of the pole position in the CE plane does not depend on the angular momentum and is sufficiently small to put  $A_2=0$ in Eq.(\ref{a2}) \cite{FH2}. Now, on the real $J$-axis for $J>>1$  we rewrite (\ref{a2}) as
 \begin{eqnarray}\label{a4}
E_1\approx A_1J^2+B_1\equiv J^2/2I+E_0\\
\nonumber
E_2\approx B_2 \equiv -1/\tau,
\end{eqnarray}
where $\tau=1/E_2$ is the lifetime of the resonance, which, we assumed, is the same whether the complex rotates or not. The first of Eqs.(\ref{a4}) is known as the '$J$-shifting approximation' \cite{JSH} and is widely used in the discussion of resonance phenomena \cite{D1,D2}. In the present context, it means that the energy of the rotating complex consists of two independent parts: its internal binding energy $E_0$ and the rotational energy of a rigid symmetric top, $J^2/2I$. This makes $I=1/2A_1$ the moment of inertia of the metastable triatomic. On the real $E$-axis, equations (\ref{a3}) now read
 \begin{eqnarray}\label{a5}
\Lambda_1 \approx J_1^2 =2I(E-E_0) \\
\nonumber
\Lambda_2\approx 2J_1J_2 =2I/\tau.
\end{eqnarray}
The first equation, which is just the inverse of the first of the Eqs.(\ref{a4}), uses the energy conservation to select the angular momentum at which the complex would be formed for a given energy $E$. The second equation tells us how far the complex would rotate during its lifetime. The typical angle of rotation is given by the life-angle $\varphi_{life}$, which is, in this case, the product of the angular velocity of the complex, $\omega\equiv J_1/I$, and the lifetime of the resonance $\tau$, $\varphi_{life}\equiv 1/J_2=\omega \tau$. 

In the above analysis we considered a subset of  $\mathcal{R}$ well removed from branching singularities, and small enough to justify using the Taylor expansion of ${\boldsymbol E}(\Lambda)$ truncated after linear terms, so that the condition defining ${\boldsymbol E}({\boldsymbol \La})$ or 
${\boldsymbol \La}({\boldsymbol E})$ can be written as
 \begin{eqnarray}\label{f11}
{\boldsymbol  F}({\boldsymbol E},{\boldsymbol \La})\equiv {\bf a}{\boldsymbol E}+{\bf b}{\boldsymbol \La}+{\bf c}=0,
\end{eqnarray}
where ${\bf a}$,${\bf b}$ and ${\bf c}$ are complex valued constants. The approximation (\ref{f11}) would break down, for example if the energies of two resonances are aligned for certain values of the centrifugal potential. This is the case for example of the F+H$_2$ reaction \cite{FH3,DARIO04} where the alignment of the resonances $\A$ and $\B$ gives specific effects in the energy dependences of the reaction probabilities \cite{TCA11} and on the angular distributions \cite{FH5} also observed experimentally \cite{Sci06}. If this happens, a branching point of the Riemann surface moves close to the real axis, and can no longer be ignored. There the function ${\bf F}({\boldsymbol E},{\boldsymbol \La})$ in (\ref{f11}) must be approximated by a polynomial quadratic in both ${\boldsymbol  E}$ and ${\boldsymbol \La}$, as discussed in details in Refs. \cite{FH3,ICS3,JCP2}. Finally, in the unlikely case of three or more resonances aligned for certain value of $J$, ${\bf F}$ must be approximated by a polynomial of the third or higher order, so that a larger number of Riemann sheets will be taken into account. 

\section{The near-threshold range:  resonances B, C,  D, and E}

The lower energy region, $ 58.54$ $meV  \le  E \le 78.54$ $meV$, is rich in resonances, and the resonance patterns in Fig.\ref{fig:1}a are more diverse (cf. Fig. \ref{fig:1}b). We will analyse each of the transitions in (\ref{b1a}) in detail after a brief remark concerning the types of Regge trajectories we are likely to encounter. 

Based on an analysis of a simple potential scattering model, in Ref. \cite{PLA} it was shown that the behaviour of a Regge trajectory in the complex ${\boldsymbol J}$-plane depends on whether at $J=0$ it correlates with a metastable state at a complex energy ${\boldsymbol E}_0=E_1+iE_2$, $E_1>0$ (trajectory of {\it type I}), or with a bound state at a negative energy $E_1<0$ (trajectory of {\it type II}). The difference between two kinds of trajectories can be understood as follows. For a sharp resonance, finding a Regge pole at an energy $E>0$ amounts to looking for a value of $J$, such that the centrifugal potential (CP) proportional to $J(J+1)$ would align the real part of the energy of the metastable state with the chosen $E$. Starting in a bound state with a negative energy, for any $E>0$ we require a positive CP capable of lifting the state to the desired level. Thus, for a trajectory of type $(II)$ the real part of ${\boldsymbol J}$ increases with $E$, while $\Imm {\boldsymbol  J}$, at first small, increases as the state is being forced out of the well. The situation is different if we start from a metastable state. For energies smaller than $E_1$, the state needs to be {\it lowered}, and that requires a {\it negative} CP or, equivalently, an imaginary ${\boldsymbol J}$. Thus, a type $(I)$ trajectory, for $E<E_1$ must first have a considerable decreasing imaginary part, while its real part should increase for $E>E_1$, where the state will, again, have to be lifted. 

This over-simplistic picture is, however, useful, and can be adapted to multi-channel scattering, currently of interest to us. Let us assume that a particular transition becomes energetically possible for $E$ above its threshold energy $E_{thr}$. For $J=0$, some of the many metastable states of the system will lie below, and some will lie above $E_{thr}$ (in terms of real parts of their energies). Those lying below will have to be 'lifted', so that their Regge trajectories will belong to the type $(II)$ and exhibit a behaviour similar to that shown in Fig. \ref{fig:4}a.
 Those above $E_{thr}$ will need to be 'lowered' initially, and their imaginary parts, will first have to decrease before, possibly, picking up later. 

\begin{figure}
\centering
\includegraphics[width=12cm,height=12cm]{{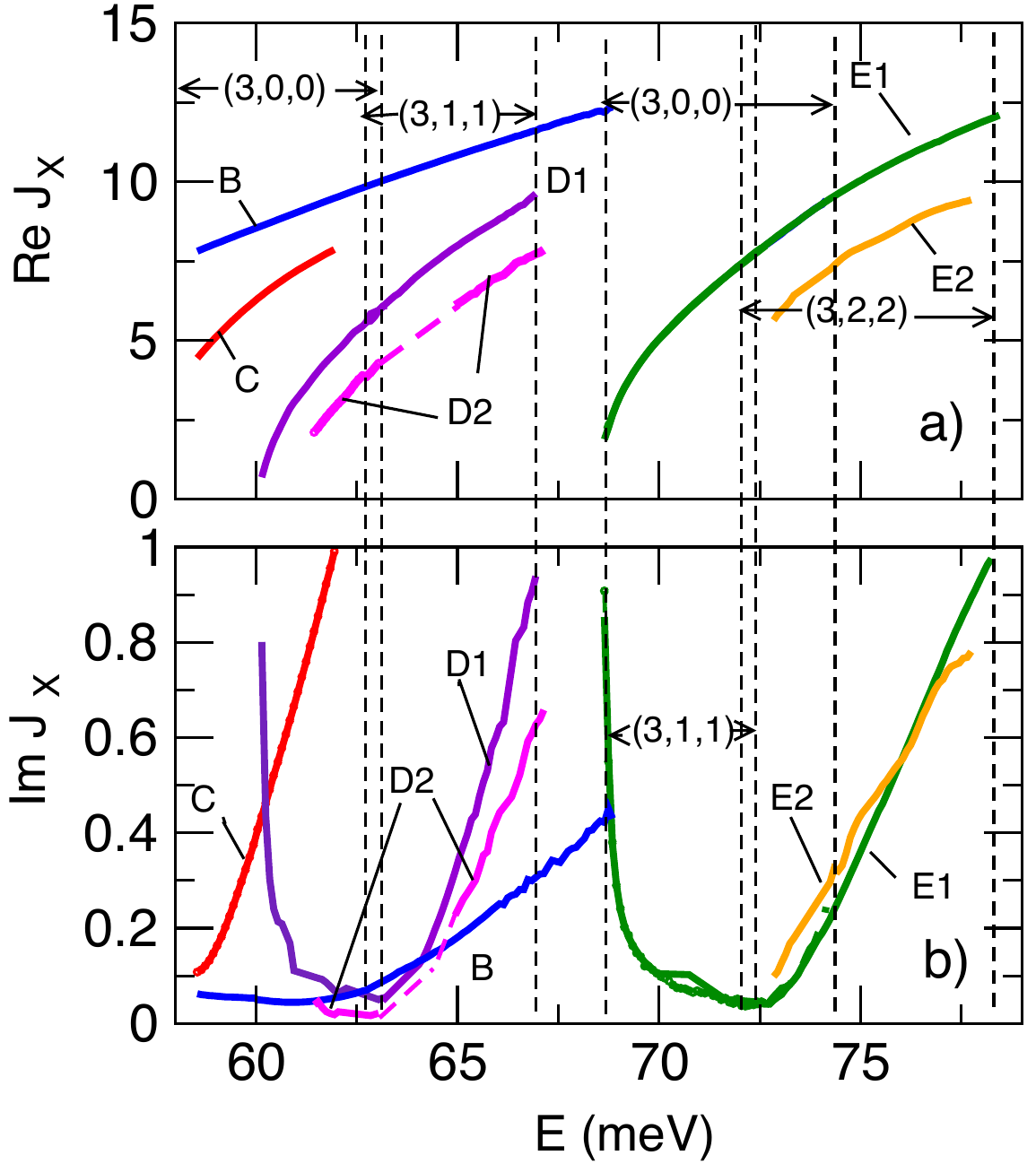}}
\caption{ a) Real parts of Regge trajectories $\B$, $\C$, $\D1$, $\D2$, $\E1$ and $\E2$ evaluated by \PD reconstruction of the ${\bf S}$-matrix elements for the transitions listed in (\ref{b1a}). Trajectory $\B$ is obtained from the $(3,0,0)\gets(0,0,0)$ transition in the whole energy range. For other trajectories, the vertical dashed lines mark the energy ranges where the part of a trajectory was found in a particular transition $(3,j',\Omega')\gets (0,0,0)$. Two parts of the $\D2$ trajectory are joined by a dashed line where it does not appear in the \PD reconstruction.
b) the imaginary parts of the trajectories shown in a). }
\label{fig:6}
\end{figure}

 The Regge trajectories obtained by \PD reconstruction of the ${\bf S}$-matrix elements for the three transitions in (\ref{b1a}) in the lower energy range are shown in Fig.\f{fig:6}. 
As one would expect (cf. Fig. \ref{fig:2}a), for the HF($v'$=3) product the Regge trajectories corresponding to resonances $\E$ and $\D$ behave as trajectories of type $(I)$, whereas those corresponding to the resonances $\B$ and $\C$ are of type $(II)$. 
 We also note  that the resonances $\D$ and $\E$ are in reality multiplets, of which only one member was found in the CE \pd reconstruction of the $(3,0,0)\gets(0,0,0)$ matrix element shown in Fig. \ref{fig:2}.
 The physical origin of the splitting of these resonances has been investigated in depth in Ref. \cite{JCP05} for the case of the F+H$_2$ reaction, where an
 inelastic adiabatic model attributes metastable states to bound levels supported by the adiabatic curves. In the space-fixed (SF) representation, one can use three quantum numbers, initial $\{v,j,l\}$ or final $\{v',j',l'\}$,
 in order to identify each component of the multiplet. The last of the quantum numbers, $l$ ($l'$) , labels the orbital angular momentum, and can take integer values from from $|J-j|$ to $J+j$, allowed by inversion parity conservation. 
  Alternatively, in the body-frame (BF), a component of the multiplet  can be labeled by $\{v',j',\Omega'\}$ quantum numbers, where $\Omega'$ stands for the projection of the hindered rotor angular momentum $j'$ along the principal axis of the BF, which correlates adiabatically with the bending motion levels of the triatomic complex. In particular, a simple hyper-spherical model \cite{JCP05} applied to the resonance $\D$, relates the amount of the splitting to the magnitudes of the Coriolis forces. In the inelastic adiabatic model, the $\D$ resonance is trapped in the adiabatic curve of $j'$=1, and it is a doublet, whereas the $\E$ resonance has three different components.

 In general, our \pd technique is able to find trajectories whose imaginary parts are not too large, and whose residues are not too small. The real parts of the poles positions are usually determined more accurately than their imaginary parts. (For someone  watching a boat from land, it is also easier to evaluate its displacement to the right or left of the observer than its distance from the shore). Because  the Coriolis effects are small for the system we study, the higher components of the multiplets (correlating with higher $\Omega'$ states) should be less pronounced in transitions from the ground rotational state, $j=0$. With smaller residues, we expect these higher component trajectories to be also harder to find with the help of our \pd technique. 

In Fig. \f{fig:6} we can distinguish six  different trajectories. The trajectory  corresponding to the resonance $\B$ is found in the $(3,0,0)\gets(0,0,0)$ and $(3,1,1)\gets(0,0,0)$ transitions in the range $58.54$ to $68.64$ $meV$. The Regge trajectory $\C$ is found only in the $(3,0,0)\gets(0,0,0)$ transition in the energy range $58.54$ to $61.94$ $meV$, where two other transitions are still closed. Both $\B$ and $\C$ trajectories belong to the type $(II)$, as expected.

The Regge trajectory corresponding to the CE trajectory $\D$ is found  in $(3,0,0)\gets(0,0,0)$ and $(3,1,1)\gets(0,0,0)$ transitions in the ranges $58.54$ to $63.14$ and $63.14$ to $67.94$ $meV$, respectively. This is a trajectory of type $(I)$, whose imaginary part shows a characteristic minimum at about $68.2$ $meV$. As expected, it appears to be a part of a doublet, so we label it $\D1$. The other member of the doublet, labelled $\D2$, is much weaker. We find the evidence of this  in the transition $(3,0,0)\gets(0,0,0)$ between $61.44$ and $63.14$ $meV$, and in the transition $(3,1,1)\gets(0,0,0)$ between $64.94$ and $67.14$ $meV$. Although we have been unable to see it for $63.13$ $meV < E < 64.94$ $meV$, there is little doubt that the two curves in Figs. \f{fig:6}a and  \f{fig:6}b belong to the same Regge trajectory. 

As in the case of the $\D$ resonance, we find two similar Regge trajectories in the energy range $68.44$ and $78.14$ $meV$, henceforth called $\E1$ and $\E2$. Trajectory $\E1$ is the Regge counterpart of the CE trajectory $\E$ in Fig. \f{fig:2} and is seen in all three transitions, contributing various parts of the curve in Figs. \f{fig:6}a and  \f{fig:6}b. Like $\D1$, it is a trajectory of type $(II)$.  We see the trajectory $\E2$ in the $(3,2,2)\gets(0,0,0)$ transition for $E$ between $72.94$ and $77.54$ $meV$. It is weaker than $\E1$ and its imaginary part can be determined less accurately.

With the list of relevant Regge trajectories complete, we proceed to identify the trajectories responsible for various patterns in the ICS's in Fig.\f{fig:1}.

\subsection{The $(3,0,0)\gets(0,0,0)$ transition}

This transition, which opens at a collision energy of $58.19$ $meV$, and covers all the lower energy range, is affected by most of the Regge trajectories shown in Fig. \f{fig:6}. Among the patterns, whose origin we need to explain, there are sharp peaks at $E \approx 59$, $61$ and $63.2$ $meV$, smoother humps at $E \approx 59.8$, and $65.5$ $meV$, and a jugged structure between $69.5$ and $74$ $meV$ (see Fig.\f{fig:7}a ). There is also a $W$-shaped feature at $E\approx 62.15$ $meV$. The results of our analysis are shown in Figs.\f{fig:7} b, c, and d.

Trajectory $\B$ contributes to the ICS the second and the third sharp peaks, together with the hump at $65.5$ $meV$, as is shown in Fig.\f{fig:7}c. All three features arise as the trajectory $\B$ passes near a real integer value $\Ree {\boldsymbol  J}_{A}=9$, $10$ and $11$. As $\Imm {\boldsymbol J}_{A}$ increases, the features become less pronounced. For comparison we have plotted by a dashed line the Breit-Wigner special case ($\Ree [\boldsymbol \gamma_n^K] =0$)  of the Fano approximation (\ref{a15}) including only the $K=9$, $10$ and $11$ terms. As expected, the approximation correctly describes the peak with the smallest $\Imm {\bf J}_{A}$, and becomes much less accurate at higher energies. Trajectory $\C$ is similar to the trajectory $\B$, and is responsible for the first peak and the first hump at $\Ree \boldsymbol J_{C}=5$ and $6$, respectively (see Fig. \f{fig:7}b). Figure Fig. \f{fig:7}d shows that trajectory $\E1$ is responsible for the saw-like structure between $69.5$ and $74$ $meV$, where its imaginary part is at minimum (cf. Fig. \f{fig:6}b). The maxima of $\sigma^{res}_{E1}(E)$ coincide with the energies at which $\Ree {\boldsymbol J}_{E1} =3,4...8$. The structure is also reasonably well described by the Fano formula (\ref{a15}), shown in Fig. \f{fig:7}d by  a dashed line.

Subtracting from $\sigma_{(3,0,0)\gets(0,0,0)}(E)$ resonance contributions from trajectories $\B$,$\C$, and $\E1$ does not yet give us a smooth structureless background (see Fig. \f{fig:7}a). There is still the $W$-shaped structure at $61.15$ $meV$. Another such structure, previously hidden by the large peak at $62.15$ $meV$, has appeared at $E\approx 63$ $meV$, together with additional dips at $60.5$, $61$ and, $61.5$ $meV$. All these features are the resonance contributions from weak resonances $\D1$ and $\D2$, whose trajectories pass close to the real axis (see Fig. \f{fig:6}b), and which only contribute when $\Ree {\boldsymbol J}_{D1,D2}$ is very close to an integer value $K$. Since the energy grid of $0.1$ $meV$ used in these calculations is relatively coarse, the contributions appears at only one energy in the vicinity of $E^K$, producing there a $V$-shaped dip. Two such dips close to each other yield the $W$-like patterns seen in Fig. \f{fig:7}a. To show that this is the case, in  Fig. \f{fig:8} we increased the energy scale and plotted the resonance contributions from the trajectories $\D1$ and $\D2$, which affect $\sigma_{(3,0,0)\gets(0,0,0)}(E)$ at the energies $E^K$ where $\Ree {\boldsymbol J}_{D1}=2,3,4,5,6$ and $\Ree {\boldsymbol J}_{D2}=3,4$. Subtracting all resonance contributions from the exact ICS now gives us a reasonably smooth background, and this concludes out analysis of the  $(3,0,0)\gets(0,0,0)$ transition.
\begin{figure}
	\centering
		\includegraphics[width=10.3cm,height=5cm]{{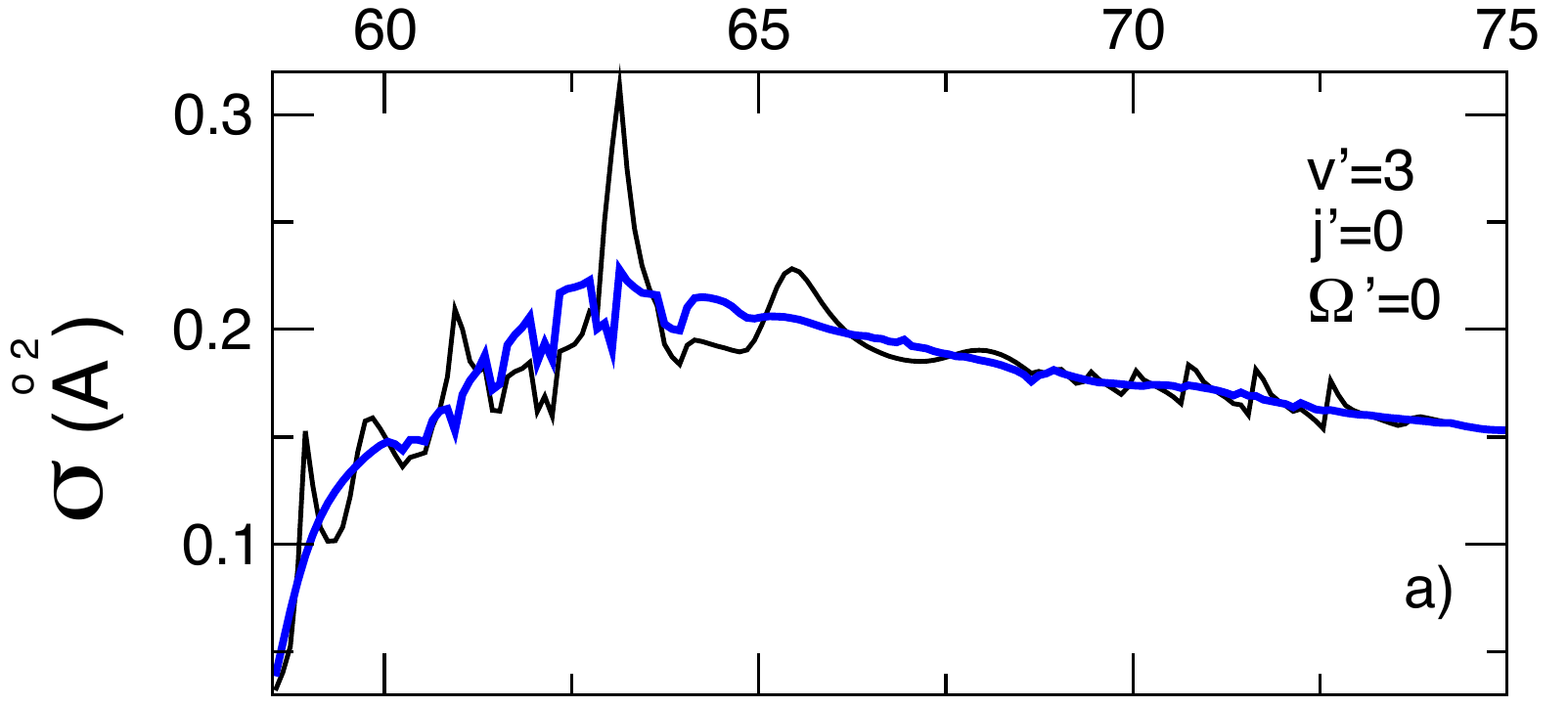}}
		\includegraphics[width=10.1cm,height=8cm]{{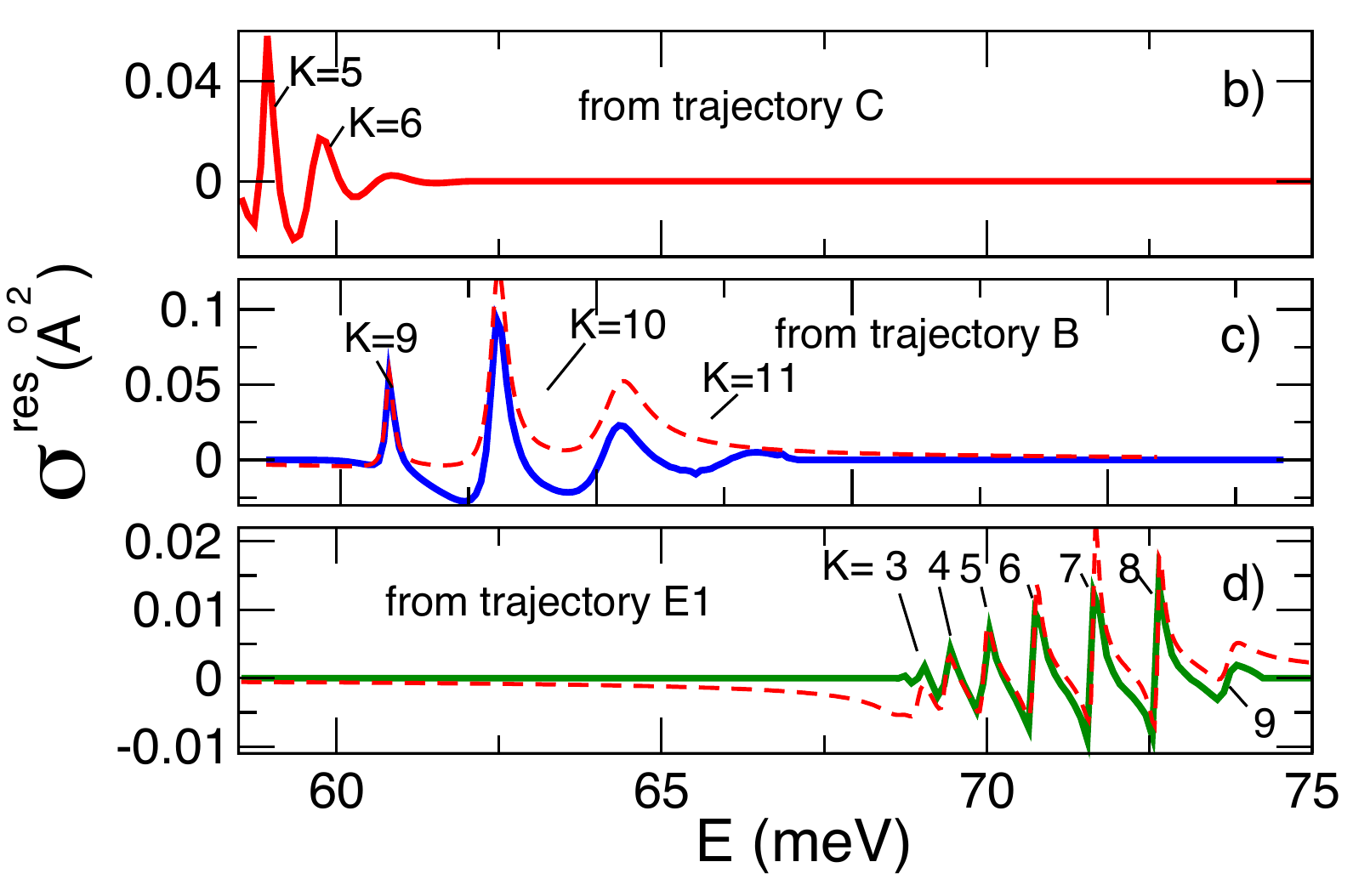}}
\caption{ a) Integral cross sections $\sigma_{\nu'\gets\nu}(E)$, $\nu=(0,0,0)$, $\nu'=(3,0,0)$, in the lower energy range (solid). Also shown by a thick solid line is $\sigma_{\nu'\gets\nu}-\sigma_B^{res}-\sigma_C^{res} -\sigma_{E1}^{res} $; b) resonance contribution from the trajectory $\C$, $\sigma_C^{res}(E)$. The maxima of $\sigma_C^{res}(E)$ appear at $E^K$ such that $\Ree {\bf J}_C(E^K)\approx K$, $K=5,6$; c) same as b) but for the trajectory $\B$. The dashed line corresponds to the Fano approximation (\ref{a15}); 
 d) same as b) but for the trajectory $\E1$.} 
\label{fig:7}
\end{figure}
\begin{figure}
	\centering
		\includegraphics[width=12cm,height=10cm]{{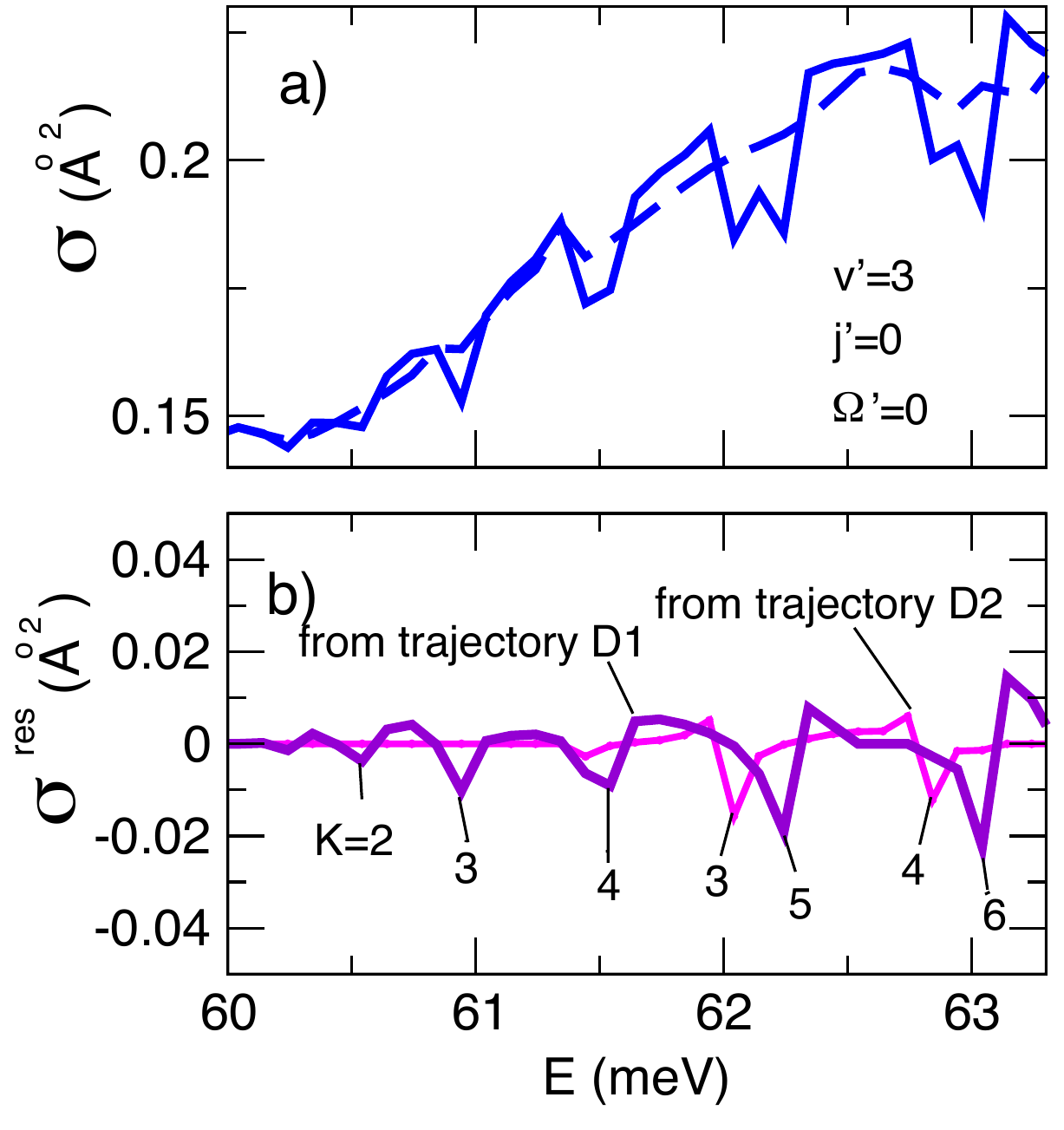}}
\caption {a) Integral cross sections $\sigma_{\nu'\gets\nu}-\sigma_B^{res}-\sigma_C^{res}-\sigma_{E1}^{res}$ (solid) and $\sigma_{\nu'\gets\nu}-\sigma_B^{res}-\sigma_C^{res} -\sigma_{E1}^{res}-\sigma_{D1}^{res}-\sigma_{D2}^{res}$ (dashed) for the same transition of Fig. \f{fig:7}. ; b) resonance contributions from the trajectories $\D1$ and $\D2$, $\sigma_{D1,D2}^{res}(E)$. The maxima of $\sigma_{D1,D2}^{res}(E)$ appear at $E^K$ such that $\Ree {\bf J}_C(E^K)\approx K$.}
\label{fig:8}
\end{figure}

\subsection{The $(3,1,1)\gets(0,0,0)$ transition}

This transition opens at $E = 62.72$ $meV$ and the corresponding ICS cannot, therefore, be affected by the resonance $\C$ (see Fig. \f{fig:6}). We see that it is also not influenced by the resonance $\B$, whose residues are very small. The two peaks and a hump in Fig. \f{fig:9}a come from the resonance $\D1$ (see Fig. \f{fig:9}b) as its Regge trajectory passes near integer values $K=6$, $7$ and $8$. The plot of $\sigma_{\nu'\gets\nu}-\sigma_{D1}^{res}$ retains small slow oscillations which can be attributed to the resonance $\D2$, but we will not follow the matter further. At slightly higher energies, the saw-like structure at $69$ $meV$ $< E <$ $73$ $meV$ comes from the resonance $\E1$, and is similar to the pattern seen in the ICS of the $(3,0,0)\gets(0,0,0)$ transition (cf. Fig. \f{fig:7}d). 

One difference is that after subtracting $\sigma_{E1}^{res}$ from the exact ICS, the resulting curve still contains a cusp at $\approx 68.7$ $meV$, as well as some minor oscillations (see Fig. \f{fig:9}c). To understand the nature of the these features, we also
plot in Fig.  \f{fig:9}c the integral term in Eq.(\ref{a14}) 
(shown by a dot-dashed line). The dot-dashed curve also contains the cusp, which must, therefore,
 be attributed to the background term $\sigma^{back}$ in Eq.(\ref{a14}). 
In Ref. \cite{PLA}, we used a simple one-channel model to argue that $\sigma^{res}$ in Eq.(\ref{a14}) does not contain the entire resonance contribution, so that a part of it remains included in $\sigma^{back}$. The cusp in Fig.\f{fig:9}a is likely to be such a feature, produced in  $\sigma^{back}$ by the resonance $\D1$.
\begin{figure}
	\centering
		\includegraphics[width=12cm,height=9cm]{{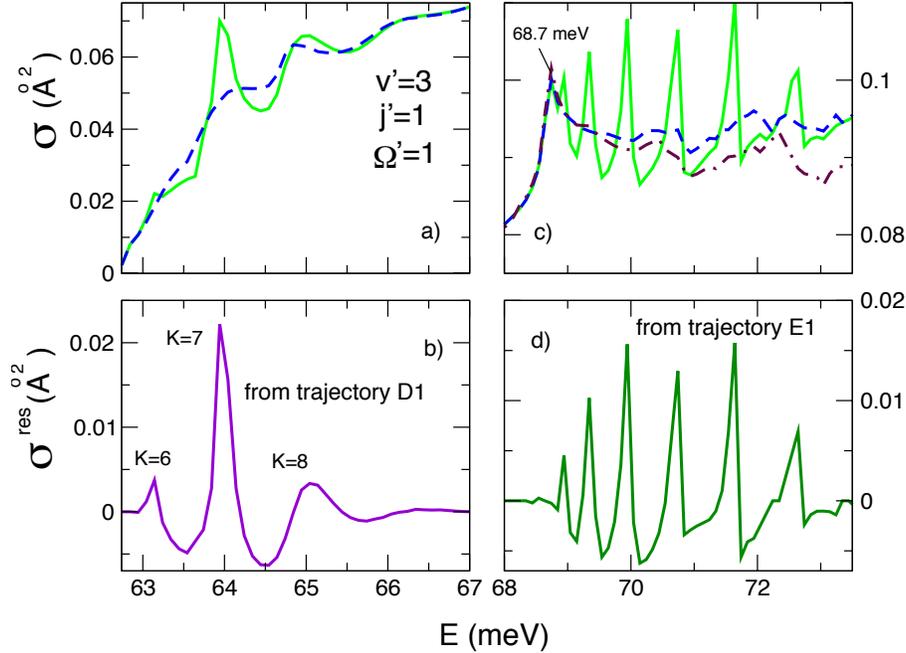}}
\caption{ 
a) Integral cross section $\sigma_{\nu'\gets\nu}(E)$, $\nu=(0,0,0)$, $\nu'=(3,1,1)$, in the lower energy range (solid). Also shown by a dashed line is $\sigma_{\nu'\gets\nu}-\sigma_{D1}^{res} $; b) the resonance contribution from the trajectory $\D1$, $\sigma_{D1}^{res}(E)$; c) the same cross sections in a different energy range. Also shown are $\sigma_{\nu'\gets\nu}-\sigma_{E1}^{res}$ (dashed), and the integral in Eq.(\ref{a14}) (dot-dashed); d) the resonance contribution from the trajectory $\E1$, $\sigma_{E1}^{res}(E)$.}
\label{fig:9}
\end{figure}

\subsection{The $(3,2,2)\gets(0,0,0)$ transition}

This transition opens at a higher energy, $E=71.77$ $meV$, and can only be affected by the resonances $\E1$ and $\E2$. The resonance $\E1$ contributes to the ICS two Lorentzian peaks, the first of which is accurately described by the formula (\ref{a15}) (see Fig. \f{fig:10}b). The peaks occur as $\Ree J_{E1}$ approaches the values $K=8$ and $9$, respectively. The contribution from the trajectory $\E2$ is about five times smaller, and consists of a Fano-like shape at $E\approx 72.05$ $meV$ and a Lorentzian peak at $E\approx 74.25$ $meV$ (see Fig. \f{fig:10}c). The smooth background cross section $\sigma^{back}$ is shown in Fig.\f{fig:10}a.
\begin{figure}
	\centering
		\includegraphics[width=9.9cm,height=6cm]{{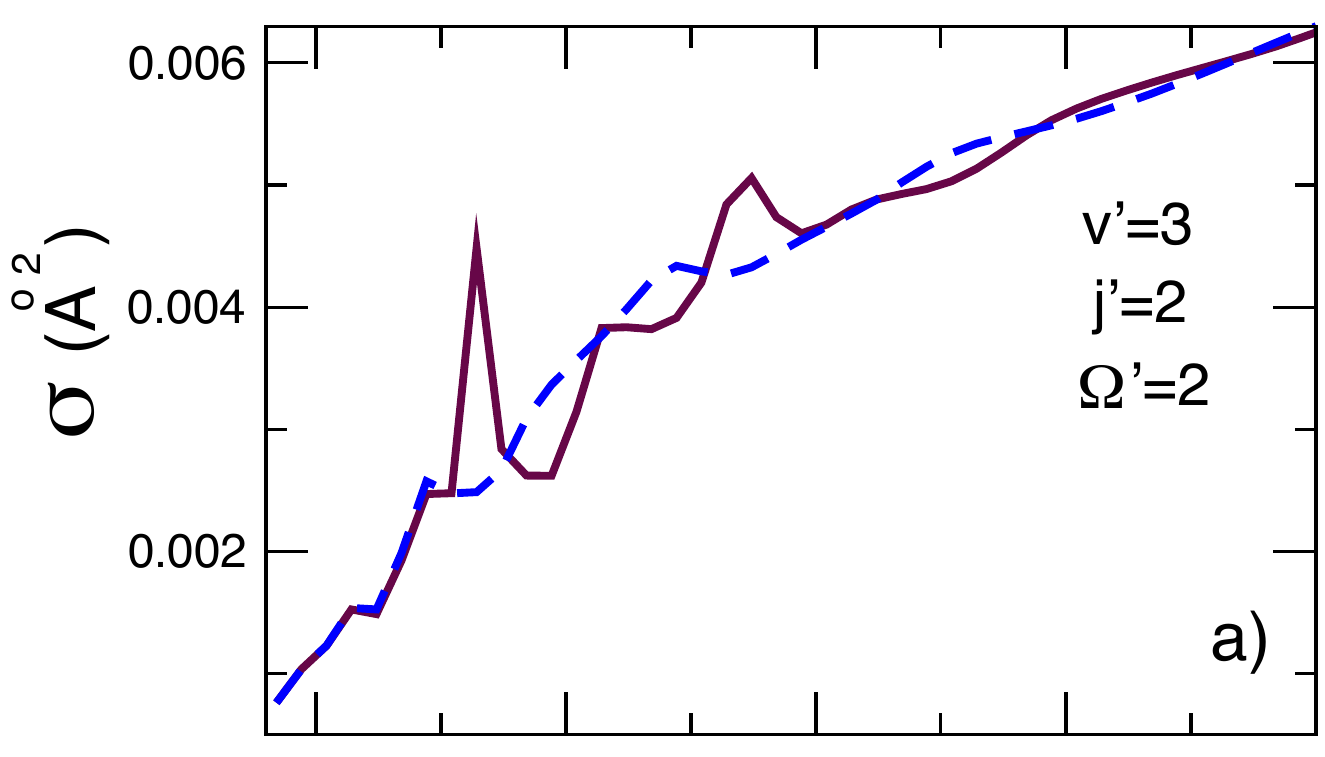}}
		\includegraphics[width=10.4cm,height=8cm]{{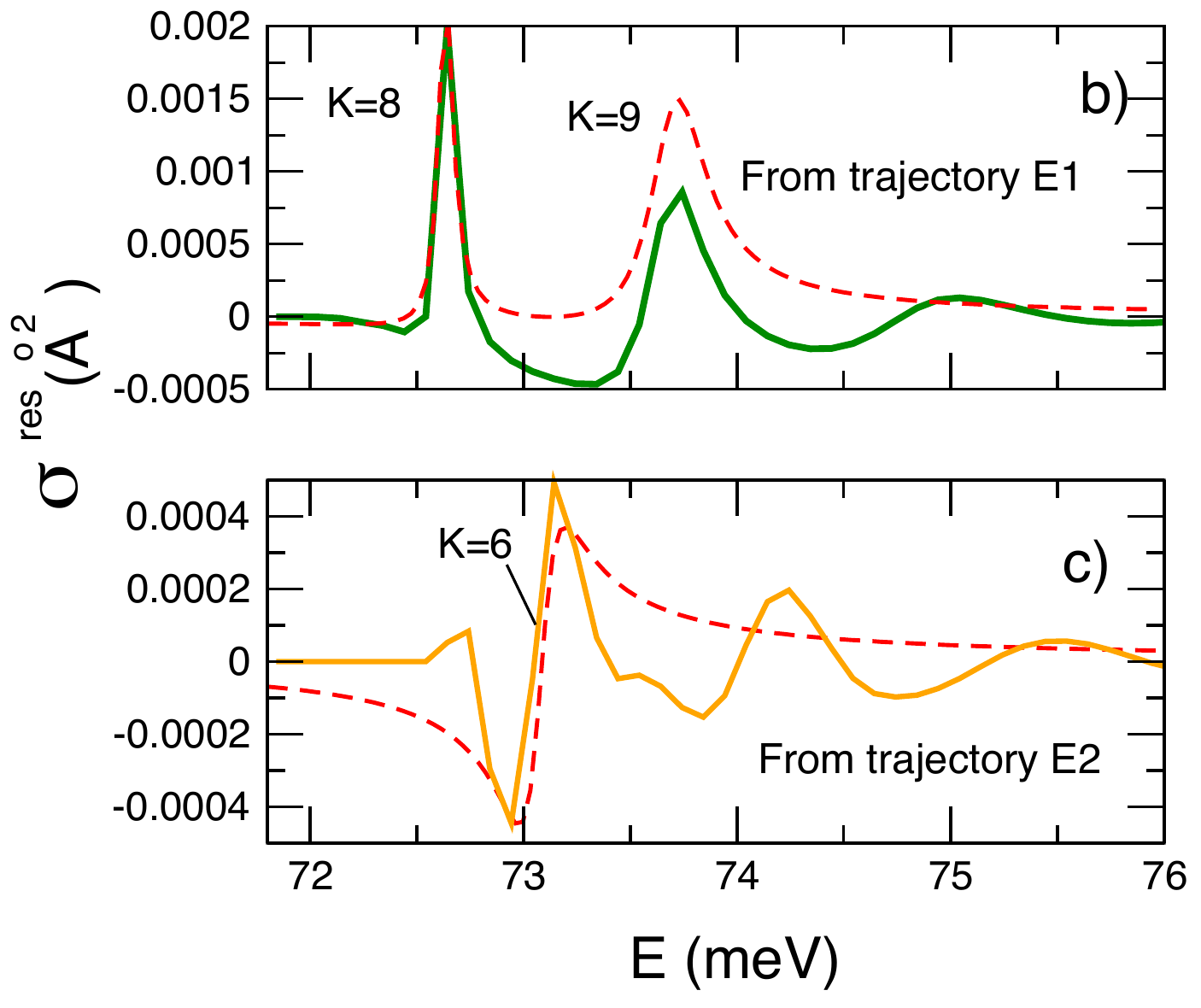}}
\caption{ a) 
Integral cross sections $\sigma_{\nu'\gets\nu}(E)$, $\nu=(0,0,0)$, $\nu'=(3,2,2)$, in the lower energy range (solid). Also shown by a dashed line is $\sigma_{\nu'\gets\nu}-\sigma_{E1}^{res}-\sigma_{E2}^{res} $; b) resonance contribution of the trajectory $\E1$, 
$\sigma_{E1}^{res}(E)$. The dashed line corresponds to the Fano approximation (\ref{a15}); c) same 
as b) but for the trajectory $\E2$.}
\label{fig:10}
\end{figure}

\section{Conclusion and discussion}

In summary, there are several resonances which affect the behaviour of the integral cross sections of the title reaction in the energy range $ 58.54$ to $197.54$ $meV$  \cite{D2}. Some have been identified earlier in \cite{D1} as one transition state resonance ($\A$) and several Feshbach resonances arising from the capture in metastable states of the van der Waals well in the exit channel ($\B$,$\C$, $\D$ and $\E$). 
In this work we present a quantitative CAM analysis of the effects produced on state-to-state ICS of the title reaction.
As the input, we use the scattering matrix elements  evaluated with the FXZ potential energy surface (PES) \cite{PES4}.
This  recent {\it ab initio} PES is known to accurately reproduce many resonance features experimentally observed for the F+HD reaction \cite{EXP6,EXP7,EXP8,EXP9}.
\newline
 Positions and widths of the metastable states affecting the state-to-sate integral cross sections, obtained by \PD reconstruction of the $S$-matrix element in the complex energy plane, are shown in Fig. \f{fig:2} for different values of the total angular momentum $J$. 
 The behaviour of the complex energy poles of the resonances $\A$,$\B$,$\C$ and $\D$ are in good agreement with those obtained by $Q$-matrix analysis in \cite{D1}, notwithstanding significant differences in the exit Van der Waals well of the PES's employed in 
 \cite{D1} and the present work (for a comparison, see Fig. 1 of Ref.\cite{D2}). 
 There is also a resonance $\E$ at somewhat higher energies, not discussed by the authors of \cite{D2}. The similarity of the resonance poles structures obtained for different PES's, 
and  the high energy resolution achieved in the {\it state of the art} molecular beam experiments \cite{EXP6,EXP7,EXP8,EXP9}, 
suggest that the resonance effects, studied in this work, should be accessible to experimental observation.
Such an experiment would provide a highly sensitive test of the accuracy of quantum chemistry calculations. 

However, the knowledge of complex energies is not enough to quantify the effects which the resonances produce in the state-to-state integral cross sections. For this purpose, we require the Regge trajectories corresponding to the resonances ${\mbox A}$,$\B$,$\C$, $\D$ and $\E$, which are shown in Figs. \f{fig:4} and \f{fig:6}.
Our analysis
demonstrates that the resonances $\D$ and $\E$ are multiplets, confirming the splitting phenomenon of bending excited metastable states studied in \cite{JCP05} for the F+H$_2$ reaction. The Regge trajectories exhibit two different types of behaviour \cite{PLA}. While their real parts tend to always increase with energy, the imaginary parts may increase (resonances $\A$, $\B$ and $\C$), or first decrease with the energy (resonances $\D1$ and $\E1$), depending on whether the position of the corresponding metastable state at $J=0$, $\Ree {\boldsymbol E}_X(J=0)$, $X=\A,\B,\C...$ lies above or below the energy threshold at which the channel opens. The explanation is similar to the one given in \cite{PLA}: lowering the state to an energy $E<\Ree {\boldsymbol E}_X(J=0)$ requires a negative centrifugal potential and, therefore, a large $\Imm {\boldsymbol J}$ (cf. Figs.\f{fig:2} and 6) . Raising the state to an energy $E>\Ree {\boldsymbol E}_X(J=0)$ requires a positive centrifugal potential. However, a large centrifugal barrier also alters the overall potential, which can no longer support long-lived metastable states, whose lifetimes and life angles must, therefore, decrease. As a result, the Regge trajectories move deeper into the complex ${\boldsymbol J}$-plane, and cease to affect the ICS's as energy increases. Accordingly, resonance effects remain predominantly a low-energy phenomenon. 

It is worth mentioning the close relation which exists between complex energies and Regge poles, so that the knowledge of a trajectory of one type may allow reconstruction of its other counterpart. We considered the simplest case of this relation in Sect. V, and used it to relate the CE and the Regge trajectories of the resonance $\A$ in the higher energy region (see Fig.\f{fig:5}).

Having identified the relevant Regge trajectories, we followed them to evaluate the contributions they make to the ICS of the chosen transition. These contributions may take various forms depending on the proximity of the trajectory to the real ${\boldsymbol J}$-axis, and the magnitude of its residue. In the higher energy range a single trajectory corresponding to the resonance $\A$ is responsible for modulated sinusoidal oscillations [see Eq.(\ref{a18})] for all transitions considered (see Fig.\f{fig:3}). The oscillations for different transitions are in phase, and the resonance pattern survives summation over helicities and rotational quantum numbers \cite{D2}. This allows these oscillations to be visible in recent molecular beams experiments \cite{EXP7}. In the lower energy range, for the $(3,0,0)\gets(0,0,0)$ transition, we assign the first two, and the subsequent three peaks, to the trajectories $\C$ and $\B$, respectively. The first peaks of the sequences have nearly Lorentzian shapes and are well described by the Fano-like formula (\ref{a15}). A similar sequence of peaks is contributed to the ICS of the $(3,1,1)\gets(0,0,0)$ transition by the trajectory $\D1$. The contribution of the trajectory $\E1$ to $\sigma_{(3,0,0)\gets(0,0,0)}(E)$ consists of seven asymmetric Fano shapes, which are produced while its imaginary part is small. A similar pattern is produced by it in $\sigma_{(3,1,1)\gets(0,0,0)}(E)$, where the Fano shapes are inverted, with initial gradual rise followed by a sharp fall. The trajectory $\E1$ also affects the $(3,2,2)\gets(0,0,0)$ transition, although in a different manner. There it produces two nearly Lorentzian peaks, followed by a low broad hump. The $(3,2,2)\gets(0,0,0)$ ICS is also affected by a much weaker $\E2$ resonance, with $\sigma^{res}_{(3,2,2)\gets(0,0,0)}(E)$ consisting of an asymmetric Fano shape followed by a Lorentzian peak.

Finally, the accuracy of a \PD reconstruction depends on the accuracy of the input data and on the numerical format (in this work, exponential with four significant digits) they are stored in. In general, the quality of the input data was found sufficient to accurately  reproduce a Regge trajectory, with its different parts consistently reproduced in the analysis of different transitions. Determination of weak (small residue) trajectories passing close to the real axis proved to be most difficult. Such are the trajectories $\D1$ and $\D2$, which make a notable contribution only at $E\approx E^K$ when ${\boldsymbol J}_X$ is close to an integer $K$, where $\Imm {\boldsymbol J}_X$ can be determined with a good accuracy. For energies between $E^K$, $\Imm {\boldsymbol J}_X$ found by \PD reconstruction fluctuates. Fortunately, this is not important, since there the contribution of such a trajectory vanishes. Thus, we can connect $\Imm {\boldsymbol J}_X(E^K)$ by a smooth line, as was done for the resonances $\D1$ and $\D2$ in Fig. \f{fig:6}).  

To conclude, we would like to emphasise that the present CAM approach is suitable for analysing state-to-state ICS for transitions with non-zero initial and final helicities $\Omega$ and $\Omega'$.  
This makes it a convenient tool for investigating resonance effects in a broad class of reactions, e.g., those involving ions and molecules,
 where recent time independent close coupling studies \cite{He12,DDF14} have found a large number of non-zero helicity states contributing to the integral cross sections.   

 \section{Acknowledgements:}
DS acknowledges support of the Basque Government (Grant No. IT-472-10), and the Ministry of Science and Innovation of Spain (Grant No. FIS2009-12773-C02-01).
EA is grateful for support  through BCAM Severo Ochoa Excellence Accreditation SEV-2013-0323 and the MINECO project MTM2013-46553-C3-1-P. 
 DDF wants to give a particular acknowledgement to the Basque Foundation for Science for an Ikerbasque Visiting Fellowship grant in the University of the Basque Country of Leioa (Bilbao), during which most of the work was done. DDF also thanks the Italian MIUR for financial support through the PRIN 2010/2011 grant N. 2010ERFKXL. The SGI/IZO-SGIker UPV/EHU and I2BASQUE are acknowledged for providing computational resources. Quantum reactive scattering calculations ware performed under the HPC-EUROPA2 project (project number:228398) with support of the European Commission - Capacities Area- Research. Some of the computational time was also supplied by CINECA (Bologna) under ISCRA project N. HP10CJX3D4. In addition, the authors are grateful to Prof. J.N.L. Connor and Prof. V. Aquilanti for useful discussions.


\begin{thebibliography}{999}
\bibitem{FANO}   
Thus, for a resonance at $E=E^K$ Fano's formula reads \cite{FANO2} ($q$=const)  $\sigma^{res}=\frac{[(E-E^K]+\Gamma q/2]^2}{(E-E^K)^2+\Gamma^2/4}$, while Eq.(\ref{a15}) gives
$\sigma^{res}_K=\frac{Re [\gamma^K](E-E^K)+Im[\gamma^K]\Gamma/2}{(E-E^K)^2+\Gamma^2/4}$.
\end{thebibliography}
\end{document}